\documentclass[aps,pra,reprint, amsmath, amssymb,superscriptaddress,nofootinbib]{revtex4-1}
 
\usepackage{bm}
\usepackage[retainorgcmds]{IEEEtrantools}
\usepackage{graphicx}
\usepackage{mathrsfs}
\usepackage{amsmath}
\usepackage{amsfonts}
\usepackage{amssymb}
\usepackage{color}
\usepackage{times,txfonts}
\usepackage{nicefrac}
\usepackage{ragged2e}
\usepackage{tikz}
\usepackage[colorlinks=true,linkcolor=blue,urlcolor=blue,citecolor=blue,pdfusetitle]{hyperref}
\usepackage{blindtext}

\newcommand{\tset}[1]{\mathcal{T}}
\newcommand{\tsetd}[1]{\mathcal{T}_d}
\newcommand{\tseto}[1]{\mathcal{T}_o}

\newcommand{\hl}[1]{{\bf \emph{#1}}}

\renewcommand{\selectlanguage}[1]{}
\begin{document}

\title{Eigenoperator approach to Schrieffer-Wolff perturbation theory and dispersive interactions}
\date{\today}
\author{Gabriel T. Landi}
\affiliation{Department of Physics and Astronomy, University of Rochester, Rochester, New York 14627, USA}
\email{glandi@ur.rochester.edu}

\begin{abstract}

Modern quantum physics is very modular: we first understand basic building blocks (``XXZ Hamiltonian'' ``Jaynes-Cummings'' etc.) and then combine them to explore novel effects. A typical example is placing known systems inside an optical cavity. 
The Schrieffer-Wolff perturbation method is particularly suited for dealing with these problems, since it casts the perturbation expansion in terms of operator corrections to a Hamiltonian, which is more intuitive than energy level corrections, as in traditional time-independent perturbation theory. 
However, the method lacks a systematic approach.
In these notes we discuss how \emph{eigenoperator decompositions}, a concept largely used in open quantum systems, can be employed to construct an intuitive and systematic formulation of Schrieffer-Wolff perturbation theory. 
To illustrate this we revisit various papers in the literature, old and new, and show how they can instead be solved using eigenoperators. 
Particular emphasis is given to perturbations that couple two systems with very different transition frequencies (highly off-resonance), leading to the so-called dispersive interactions. 

\end{abstract}

\maketitle{}

\tableofcontents

\section{Introduction}\label{sec:intro}

The use of time-independent perturbation theory in its current format dates back to Lord Rayleigh in 1877 and the study of vibrating strings~\cite{rayleigh1969}.
In quantum mechanics, its importance was recognized very early on by Schr\"odinger, who used it to calculate the Stark shift of the hydrogen atom~\cite{schrodinger1926} in his third 1926 paper ``\emph{Quantization as a Problem of Proper Values (Part III).}''
The fact that his calculations matched with experimental observations was, to Schr\"odinger, a strong confirmation of the predictive power and flexibility of his equation: to obtain new physics one  simply needs to add new terms to his equation, and use perturbative methods to obtain the answers.
This is a paradigm that is still prevalent in many areas of physics. 

Traditionally, time-independent perturbation theory provides corrections to energy levels and eigenvectors. 
For example, the canonical textbook formula for non-degenerate perturbation theory reads 
\begin{equation}\label{energy_corrections}
    \Delta E_\mu = V_{\mu\mu} + \sum_{\nu\neq \mu} \frac{|V_{\mu\nu}|^2}{E_\mu-E_\nu},
\end{equation}
where $E_\mu$ are the unperturbed energies and $V_{\mu\nu}$ are the matrix elements of the perturbation in the unperturbed eigenbasis. 
Extensions to degenerate spectra are commonly taught and applied. 
In textbooks, these formulas are applied to famous problems, such as the Zeeman and Stark shifts, or the fine and hyperfine corrections to atomic spectra. 

Nowadays, however, thinking in terms of energy corrections, as in~\eqref{energy_corrections}, has become less convenient. 
Instead, we are trained to think in terms of \emph{Hamiltonians.}
The reason is because quantum physics have become much more modular: 
we first learn how to think in terms of elementary Hamiltonian blocks:  ``Jaynes-Cummings,'' ``Rabi,'' ``tight-binding,'' ``XXZ,'' ``Hubbard,'' ``Dicke,'' etc.
And then we study what happens when we combine them in different ways.
A common example is confining a system inside an optical cavity. 
In fact, presently it seems very fashionable to put all sorts of weird things inside optical cavities. 
Countless more examples could also be mentioned, in all types of quantum coherent platforms (for an extreme example, see~\cite{lee2022}).

This modularity strongly motivates the need for a time-independent perturbation theory that is phrased in terms of Hamiltonians, instead of energy corrections.
Fortunately, such a theory already exists, and is called the Schrieffer-Wolff (SW) transformation~\cite{schrieffer1966}.
Historically, similar methods had already been used before, by Van Vleck~\cite{vanvleck1929}, L\"odwin~\cite{lowdin1962}, Fr\"olich~\cite{frohlich1952}, and Foldy and Wouthuysen~\cite{foldy1950}. 
But Schrieffer and Wolff popularized it in the many-body physics context, as a way to relate the Anderson and Kondo models. 
This has since led to strong interest from the condensed matter community in deriving effective low energy Hamiltonians for many-body systems~\cite{bravyi2011,datta2011,bukov2016,haq2020,kale2022,day2024,dag2023,hu2023}.
The method also appears from time to time in the atomic, molecular and optical physics community~\cite{CohenTannoudji1998,gerry,krantz2019,mercurio2024}. 
And in quantum information sciences, it picked up steam with the derivation of the dispersive Jaynes-Cummings interaction, which plays an important role in superconducting devices~\cite{blais2004,wallraff2004,koch2007,boissonneault2009,blais2021}. 
Various other applications have also been investigated, including extensions to open dynamics~\cite{kessler2012,sciolla2015,jager2022,malekakhlagh2022,vanhoecke2024}, periodically driven systems~\cite{bukov2016,wang2024a}, non-Hermitian physics~\cite{starkov2023}, as well as connections with quantum algorithms~\cite{wurtz2020,zhang2022,haq2022,mozgunov2023,marecat2023} and information geometry~\cite{pinter2024,kowalski2022}.
Even a numerical library has recently become available~\cite{day2024}.

\begin{figure*}[!th]
    \centering
    \includegraphics[width=0.9\linewidth]{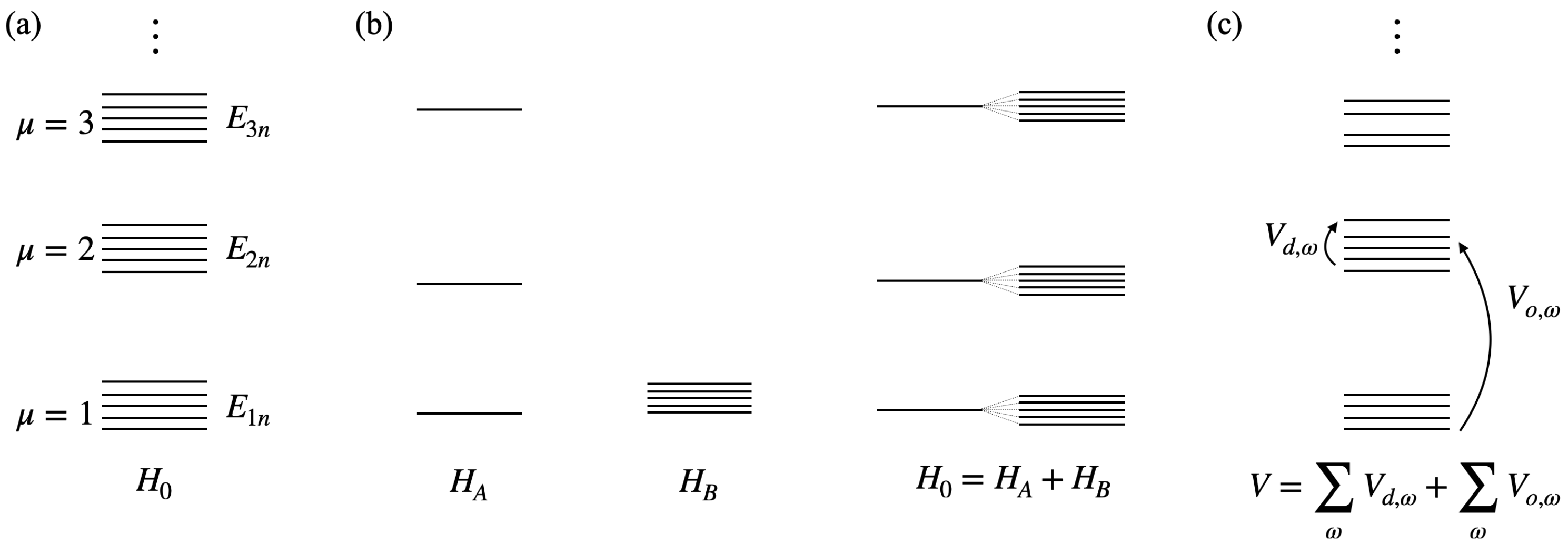}
    \caption{(a) The Schrieffer-Wolff (SW) perturbation theory applies to systems where the energy levels are roughly bundled into blocks in energy space. Each $\mu$ represents a subspace, and the energies $E_{\mu n}$ within that subspace need not be exactly equal, but are assumed to be close. 
    (b) A typical scenario where the SW method applies is when the perturbation $V$ couples two systems, $A$ and $B$, which have very different energy level spacings (the so-called dispersive regime).
    (c) A perturbation can always be decomposed in terms of eigenoperators $V_\omega$ that produce transitions with specific frequencies, either within a subspace $\omega_d \in \mathcal{T}_d$, or between subspaces, $\omega_o \in \mathcal{T}_o$. 
    To second order, the effective Hamiltonian [Eq.~\eqref{Hprime_eigen_final_d}] can be written in terms of virtual transitions $[V_{\omega_o},V_{\omega_o'}]$ that take the system away from a subspace and then back ($\omega_o-\omega_o' \in \mathcal{T}_d$).
    }
    \label{fig:drawing}
\end{figure*}

However, despite these successes, Schrieffer-Wolff remains somewhat of a niche method, which most researchers have just heard of vaguely. 
For instance, it is largely unknown that SW actually contains \emph{all} textbook time-independent perturbation theory results as particular cases, both degenerate and non-degenerate. 
From an educational perspective, one could therefore even argue that SW could completely replace many textbook treatments.

The largest difficulty with the SW method is how to apply it. 
There are two ways.
One involves solving a certain operator equation, which requires some heuristics and reverse engineering, and is therefore not very procedural. 
This is how Schrieffer and Wolff did it in their original paper~\cite{schrieffer1966}. 
The other approach is to use a cumbersome sum over many matrix elements. This is more procedural, but looses the operator character of the theory (i.e., we go back to energy levels).

Here we show that the Schrieffer-Wolff transformation can be implemented in a systematic fashion using the concept of \emph{eigenoperator decomposition}.
The term ``eigenoperator'' is not standard in physics or mathematics.
Most researchers who know the term (myself included), learned it from page 133 of the Breuer and Petruccione textbook on open quantum systems~\cite{breuer2007} (there is also a brief passage in the book of Gardiner and Zoller~\cite{gardiner2004}).
A side goal of this paper will be to highlight the usefulness of this concept in general. 
The final result will be given by Eqs~\eqref{Hprime_eigen_final_d}, which writes the effective SW Hamiltonian as a sum over eigenoperators of the perturbation. 
The results in this paper are not necessarily novel. 
But it is the hope that with eigenoperator decompositions, the method can be made systematic and intuitive, therefore opening up the way for new applications.
To corroborate this, we will rederive several results in the literature, from both the condensed matter and the AMO/QIS side. 


\subsection{Summary of the results}

Given a system with Hamiltonian $H = H_0 + V$, the SW method constructs a rotated Hamiltonian $H' = e^S H e^{-S}$ which is meant to be block diagonal in the eigenbasis of $H_0$. 
The basic recipe, to second order, is to choose $S$ as the solution of $[H_0,S] = V$. This then leads to the approximate Hamiltonian~\cite{schrieffer1966}
\begin{equation}
    H' \simeq H_0 + \tfrac{1}{2}[S,V].
\end{equation}
This method is reviewed in Sec.~\ref{sec:SW} and Appendix~\ref{app:SW_expansion}, where several subtleties skipped here are discussed. 

The main result of this paper is as follows. We use the fact that a perturbation $V$ can always be decomposed in eigenoperators of $H_0$, as~\cite{breuer2007}
\begin{equation}
    V = \sum_{\omega> 0} \big(V_\omega^{} + V_\omega^\dagger\big), 
\end{equation}
where $\omega$ are all possible transition frequencies of $H_0$.
Each term $V_\omega$ satisfies $[H_0, V_\omega] = -\omega V_\omega$ and therefore acts as a lowering operator for $H_0$.
Finding this eigenoperator decomposition is generally quite easy, and several techniques will be discussed in Sec.~\ref{sec:eigenoperator_decomposition}. 
Once this decomposition is found, the basis transformation operator $S$ can be readily written as 
\begin{equation}
    S = - \sum_{\omega> 0}\frac{1}{\omega}\big(V_\omega^{} - V_\omega^\dagger\big).
\end{equation}
And the rotated Hamiltonian becomes 
\begin{equation}
    H' \simeq H_0  - \tfrac{1}{2}\sum_{\omega,\omega'> 0} \left( \tfrac{1}{\omega}+\tfrac{1}{\omega'}\right) [V_\omega^{}, V_{\omega'}^\dagger].
\end{equation}
The SW Hamiltonian is therefore described in terms of virtual transitions which take the system away from an energy subspace by an amount $\omega$, and then back by an amount $\omega'$. 
These results are discussed at length in Sec.~\ref{sec:main}, where we also show that, very often, some terms in the double sum can be dropped. 
For example, in various models the only surviving term is $\omega' = \omega$. 
The entire effort boils down to finding the eigenoperators. And the techniques of Sec.~\ref{sec:eigenoperator_decomposition} greatly facilitate that task.

Of particular interest is the case of a perturbation that couples two systems with Hamiltonians $H_A$ and $H_B$ having very different transition frequencies, $|\omega_A| \gg |\omega_B|$
(c.f.~Fig.~\ref{fig:drawing}(b)).
That is, two systems that are very far off resonance. 
The eigenoperators in this case are written as $V_{\omega_A-\omega_B}$, and finding them is as easy as in the previous case. 
The SW Hamiltonian then becomes 
\begin{equation}
    H' \simeq H_0  - \tfrac{1}{2}\sum_{\omega_A>0}\sum_{\omega_B,\omega_B'} \left( \tfrac{1}{\omega_A-\omega_B}+\tfrac{1}{\omega_A-\omega_B'}\right) [V_{\omega_A-\omega_B}^{}, V_{\omega_A-\omega_B'}^\dagger].
\end{equation}
The sum is now only over positive frequencies of system $A$, but is unrestricted on system $B$. This leads to the so-called dispersive interactions, and will be discussed in detail in Sec.~\ref{sec:dispersive}.

A list of all models studied in this paper is given in Table~\ref{tab:models}

\begin{table*}[!t]
    \centering
    \caption{Examples studied in this paper using eigenoperators. Refer to the corresponding section for the precise definition of each term. }
    \setlength{\tabcolsep}{1mm} 
    \begin{tabular}{c|c|c|c}    
         Label & $H_0$ & $V$ & Section \\[0.2cm] \hline
         &&&\\[-0.1cm]
    Qutrit & 
         $H_0 = \sum_{j=0}^2 E_j |j\rangle\langle j|$ & 
         $V = \sum_{j=0}^1 v_{j,j+1} |j\rangle\langle j+1| + \text{h.c.}$ &
         Sec.~\ref{sec:qutrit_example}.
        \\[0.3cm]\hline &&&\\[-0.1cm]
    3 boson model~\cite{mercurio2024} & 
        $H_0 = \omega_a a^\dagger a + \omega_b b^\dagger b + \omega_c c^\dagger c$ & 
        $V = - \frac{g}{2}\Big[(a+a^\dagger)^2 -\Omega^2 (c+c^\dagger)^2\Big](b+b^\dagger)$ 
        & 
        Sec.~\ref{sec:procedure_eigenops}.
        \\[0.3cm]\hline &&&\\[-0.1cm]
    1 boson model & 
        $H_0 = \omega_r a^\dagger a$ & 
        $V = g (a+a^\dagger)^n$ & 
        Sec.~\ref{sec:single_boson}.
        \\[0.3cm]\hline &&&\\[-0.1cm]
    2 boson model & 
        $H_0 = \omega_a a^\dagger a + \omega_b b^\dagger b$& 
        $V = g (a+a^\dagger)(b+b^\dagger)$& 
        Sec.~\ref{sec:two_bosons}.
        \\[0.3cm]\hline &&&\\[-0.1cm]
    Original SW calculation~\cite{schrieffer1966} & 
        $ H_0 = \sum_{k,s} \epsilon_k n_{ks} + \sum_s \epsilon_d n_{ds} + U n_{d\uparrow} n_{d\downarrow}$ &
        $V = \sum_{k,s} \big\{ V_{kd} c_{ks}^\dagger c_{ds}^{} + V_{kd}^* c_{ds}^\dagger c_{ks}^{}\big\}$ &
        Sec.~\ref{sec:SW_original}.
        \\[0.3cm]\hline &&&\\[-0.1cm]
    Dispersive boson(s) (sys. B arb.) & 
        $H_0 = \omega_r a^\dagger a + H_B$ with $\omega_r \gg |\omega_B|$ &
        $V = a^\dagger B + B^\dagger a$ (for $B$ arb.). &
        Sec.~\ref{sec:dispersive_single_bosonic}.
        \\[0.3cm]\hline &&&\\[-0.1cm]
    Jaynes-Cummings/Rabi & 
        $H_0 = \omega_r a^\dagger a  + \tfrac{\omega_q}{2}\sigma_z$ &
        $V = g(a+a^\dagger)(\sigma_++\sigma_-)$ &
        Sec.~\ref{sec:dispersive_jaynes_cummings}.
        \\[0.3cm]\hline &&&\\[-0.1cm]
    Dispersive qubit (sys. B arb.)& 
        $H_0 = \tfrac{\omega_r}{2} \sigma_z + H_B$ with $\omega_r \gg |\omega_B|$ &
        $V = \sigma_+ B + B^\dagger \sigma_-$ (for $B$ arb.). &
        Sec.~\ref{sec:dispersive_single_qubit}.        
        \\[0.3cm]\hline &&&\\[-0.1cm]
    Giant atom~\cite{roccati2024}& 
        $H_0 = \tfrac{\omega_r}{2} \sigma_z + \sum_k \epsilon_k^{} b_k^\dagger b_k^{}$  &
        $V = \sum_k g_k\big(b_k^\dagger \sigma_- + \sigma_+ b_k^{}\big)$ &
        Sec.~\ref{sec:dispersive_tight_binding_qubit}.                
        \\[0.3cm]\hline &&&\\[-0.1cm]
    Two tight-binding chains &
        $H_0 = \sum_i E_i^{} a_i^\dagger a_i^{} + \sum_k \epsilon_k^{} b_k^\dagger b_k^{}$ & 
        $V = \sum_{i,k}\Big(v_{ik}^{} b_k^\dagger a_i+ v_{ik}^* a_i^\dagger b_k^{}\Big)$ & 
        Sec.~\ref{sec:tight_binding}.
        \\[0.3cm]\hline &&&\\[-0.1cm]
    Cubic fermion-boson coupling &
        $H_0 = \sum_{i} \omega_{ri}a_i^\dagger a_i + \sum_{n,m} h_{nm} b_n^\dagger b_m^{}$ &
        $V = \sum_{i,n,m} b_n^\dagger b_m^{} \Big(M_{inm}^{} a_i^\dagger + M_{imn}^* a_i^{}\Big)$ & 
        Sec.~\ref{sec:quadratic_fermions_linear_bosons}        
        \\[0.3cm]\hline &&&\\[-0.1cm]
    Electron-phonon~\cite{frohlich1952} &
        $H_0 = \sum_{\bm{k},\sigma} \epsilon_k^{} c_{\bm{k}\sigma}^\dagger c_{\bm{k}\sigma} + \sum_{\bm{p}} \omega_{\bm{p}}^{} a_{\bm{p}}^\dagger a_{\bm{p}}$  &
        $V = \sum_{\bm{k}, \bm{p},\sigma} g_{\bm{p}} c_{\bm{k}+\bm{p}}^\dagger c_{\bm{k}}^{}(a_{\bm{p}}^{} + a_{-\bm{p}}^\dagger)$ & 
        Sec.~\ref{sec:electron_phonon}.        
        \\[0.3cm]\hline &&&\\[-0.1cm]
    Graphene in a cavity~\cite{dag2023} &
        $H_0 =  \sum_{i=L,R} \omega_{ri} a_i^\dagger a_i^{} + v_F\sum_{\bm{k}}\Big\{(k_x+ik_y) c_{A\bm{k}}^\dagger c_{B\bm{k}}^{} $  &
        $V =  -v_F \sum_{\bm{k}} \Big\{ g_L \Big(c_{A\bm{k}}^\dagger c_{B\bm{k}}^{}a_L + c_{B\bm{k}}^\dagger c_{A\bm{k}}^{}a_L^\dagger\Big) $ & 
        Sec.~\ref{sec:graphene} .              
        \\[0.2cm]
        & $\qquad+ (k_x-ik_y) c_{B\bm{k}}^\dagger c_{A\bm{k}}^{} \Big\}$ & $\qquad + g_R \Big(c_{A\bm{k}}^\dagger c_{B\bm{k}}^{}a_R^\dagger + c_{B\bm{k}}^\dagger c_{A\bm{k}}^{}a_R\Big)\Big\}$& 
        \\[0.3cm]\hline &&&\\[-0.1cm]
    Non-rel. Dirac equation~\cite{foldy1950} &
        $H_0 = \beta m $  &
        $V = - e \phi + \bm{\alpha} \cdot (\bm{p}- e \bm{A})$ & 
        Sec.~\ref{sec:Dirac}.                
        \\[0.3cm]\hline &&&\\[-0.1cm]
    $t/U$ expansion~\cite{macdonald1988,fazekas} &
        $H_0 =U \sum_i n_{i\uparrow} n_{i\downarrow} $  &
        $V = -t \sum_{\langle i,j\rangle}  \sum_\sigma  c_{i\sigma}^\dagger c_{j\sigma}^{}$ & 
        Sec.~\ref{sec:tU_expansion}.
    \end{tabular}
    \label{tab:models}
\end{table*}    
    
\section{The Schrieffer-Wolff transformation}\label{sec:SW}

\begin{figure*}
    \centering
    \includegraphics[width=0.9\linewidth]{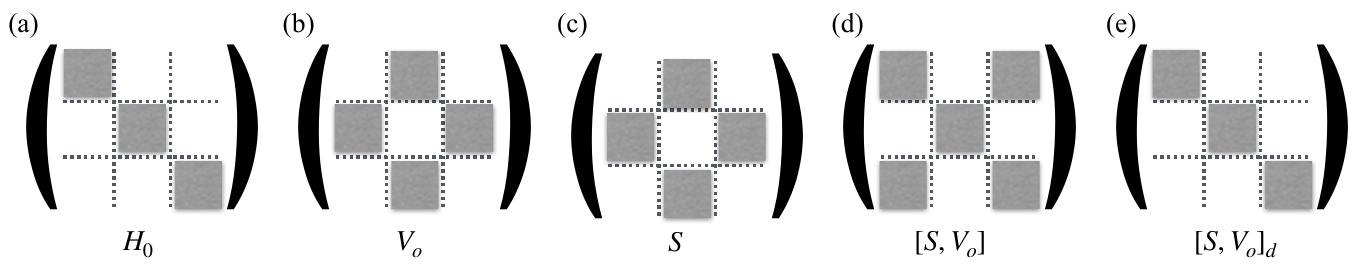}
    \caption{The block shapes of all matrices involved in the Schrieffer-Wolff transformation.}
    \label{fig:matrix_blocks}
\end{figure*}

We consider the standard scenario of time-independent perturbation theory, consisting of a system with Hamiltonian $H = H_0 +  V$. 
We know how to handle $H_0$, but wish to treat $V$ perturbatively under the assumption that it is small. 
We assume that the eigenvalues of $H_0$  are roughly bundled into blocks in energy space, as in Fig.~\ref{fig:drawing}(a).
To make this explicit, we write the eigenvalues and eigenvectors as
\begin{equation}\label{H0_eigen}
    H_0 |\mu n\rangle = E_{\mu n} |\mu n\rangle,
\end{equation}
where $\mu$ labels the different subspaces, while $n$ labels the states within each subspace. 
The definition of the $\mu$ subspaces is intentionally left somewhat loose.
But the basic idea is that $|E_{\mu n} - E_{\mu m}|$ are small within the same subspace (same $\mu$), while $|E_{\mu n} - E_{\nu m}|$ are large between different subspaces ($\nu \neq \mu$).
These ideas encompass familiar scenarios as particular cases:
\begin{itemize}\itemsep-0.1cm
    \item If a subspace $\mu$ is one-dimensional we recover non-degenerate perturbation theory; 
    \item If all states in a subspace $\mu$ have the same energy, $E_{\mu n} = E_\mu$, we recover degenerate perturbation theory. 
\end{itemize}
But the setting in Fig.~\ref{fig:drawing}(a) is more general since it allows us to consider quasidegenerate subspaces; i.e., whose energies are not exactly equal, but only bundled in energy space.  

The perturbation $V$ will, in general, connect various different energy levels. 
However, if $V$ is small, its ability to generate transitions between subspaces with very different energies will be suppressed with increasing energy separation. 
We can therefore split $V$ into a block diagonal term ($V_d$), which connects energy levels in the same subspace, and a block off-diagonal term ($V_o$) that produces transitions between subspaces.
To accomplish that, we define projectors onto each subspace,
\begin{equation}
    P_\mu = \sum_{n\in \mu} |\mu n\rangle\langle \mu n|. 
\end{equation}
Using $1 = \sum_\mu P_\mu$ we can then decompose
\begin{equation}\label{V_split}
    V 
    =  \sum_\mu P_\mu V P_\mu + \sum_{\mu\neq \nu} P_\mu V P_\nu := V_d + V_o.
\end{equation}
The total Hamiltonian is thus
\begin{equation}\label{39817198719871987391}
    H = H_0 +  V_d +  V_o.
\end{equation}
The part $H_0 +  V_d$ is block diagonal in the energy basis, but $H$ itself will generally be a full matrix because of $V_o$.

\emph{The basic idea of the SW transformation is to rotate to a basis where $H$ is block diagonal.}
That is, we define a unitary transformation
\begin{equation}\label{Hprime_def}
    H' = e^{S} H e^{-S}, \qquad S^\dagger = -S
\end{equation}
with $S$ chosen so that $H'$ is block diagonal in the $\mu$ subspaces. 
The procedure for finding $S$ is reviewed in Appendix~\ref{app:SW_expansion}. 
To second order, the basic recipe is to take $S$ as the block off-diagonal solution of 
\begin{equation}\label{S_equation}
    [H_0,S] = V_o.
\end{equation}
The solution of this equation is not unique, since we can always add a term $S\to S+f(H_0)$, for any function of $H_0$, and this would still be a solution. 
However, if we focus only on the block off-diagonal solution ($P_\mu S P_\mu = 0$), then it will be unique. 
The overall shape of the matrices $H_0,V_o,S$ is shown in Fig.~\ref{fig:matrix_blocks}(a)-(c).
With this transformation, Eq.~\eqref{Hprime_def} is approximated to 
\begin{equation}\label{Hprime_final_not_d}
    H' \simeq H_0 +  V_d + \tfrac{1}{2} [S,V_o]. 
\end{equation}
The last term captures the leading order effect of $V_o$ on each subspace. 
Higher order corrections are provided in Appendix~\ref{app:SW_expansion}.

An important subtlety, which is often overlooked, is that $[S,V_o]$ might not be block diagonal, which can occur whenever the number of subspaces is larger or equal to 3. 
For example, if $V_o$ connects subspaces $1 \leftrightarrow 2$ and $2\leftrightarrow 3$, the product $[S,V_o]$ will have both  diagonal contributions, such as $1\to 2 \to 1$, as well as  off-diagonal contributions, such as  $1\to 2 \to 3$.
This is illustrated in Fig.~\ref{fig:matrix_blocks}(d). 
If one wishes to obtain a proper block diagonal Hamiltonian, up to second order, one should discard these off-diagonal terms, and use only 
\begin{equation}\label{Hprime_final_d}
    H' \simeq H_0 +  V_d + \tfrac{1}{2} [S,V_o]_d, 
\end{equation}
where $[S,V_o]_d = \sum_\mu P_\mu [S,V_o] P_\mu$ is the block diagonal part of $[S,V_o]$. The resulting Hamiltonian will then be properly block diagonal, as in Fig.~\ref{fig:matrix_blocks}(e).
This subtlety plays an important role in deciding which terms should be kept in a series expansion, and will be illustrated below through several examples. 
As a general procedure, however, one can always start with Eq.~\eqref{Hprime_final_not_d}. 
The step from~\eqref{Hprime_final_not_d} to~\eqref{Hprime_final_d} simply amounts to throwing away terms that connect distant blocks.

To summarize, the procedure for applying the SW transformation is:
\begin{enumerate}\itemsep-0.1cm
    \item Decide how you want to bundle the energies of $H_0$ into subspaces; i.e. construct the projectors $P_\mu$;
    \item Decompose $V = V_d + V_o$ [Eq.~\eqref{V_split}];
    \item Solve Eq.~\eqref{S_equation} for $S$ (which has to be block off-diagonal);
    \item Compute $[S,V_o]$ and take its block diagonal part $[S,V_o]_d$.
\end{enumerate}
The choice of subspace projectors $P_\mu$ is entirely arbitrary. But different choices will lead to better or worse approximations. 
We also mention that in Eq.~\eqref{39817198719871987391} we could have incorporated $V_d$ into $H_0$. This could lead to a more precise approximation, but may also be more difficult to treat.

\subsection{Explicit formulas and recovering textbook results}

It is possible to write down formulas for $S$ and $H'$ in terms of the eigenvectors $|\mu n\rangle$ of $H_0$. 
Although not the focus of this paper, we list these here because of their connection to textbook treatments of time-independent perturbation theory. 
The block off-diagonal solution of Eq.~\eqref{S_equation} is
\begin{equation}\label{S_sol_elements}
    \langle \mu n| S |\nu m\rangle =
    \begin{cases}\frac{\langle \mu n| V_o| \nu m\rangle}{E_{\mu n} - E_{\nu m}}, & \nu \neq \mu,
    \\[0.2cm]
    0 & \nu = \mu.  
    \end{cases}
\end{equation}
Plugging this in Eq.~\eqref{Hprime_final_d} then yields~\cite{CohenTannoudji1998,blais2021}
\begin{equation}\label{Hprime_matrix_elements}
\begin{aligned}
    &\langle \mu n | H' |\mu  m \rangle = E_{\mu n} \delta_{n,m} + \langle \mu n | V | \mu m\rangle  \\[0.2cm]
    &+ \frac{1}{2} \sum_{\nu \neq \mu} \sum_{j\in \nu} \langle \mu n| V|\nu j\rangle \langle \nu j| V |\mu m\rangle
    \left(
    \frac{1}{E_{\mu n} - E_{\nu j}} + 
    \frac{1}{E_{\mu m} - E_{\nu j}}
    \right),
\end{aligned}
\end{equation}
and $\langle \mu n | H' |\nu m \rangle = 0$ for $\nu \neq \mu$.

It is noteworthy that the SW transformation contain all usual textbook results as particular cases. 
First, if all subspaces are one-dimensional (i.e.~there is only one level $|\mu\rangle$ in each subspace), Eq.~\eqref{Hprime_matrix_elements} reduces to 
\begin{equation}
    \langle \mu | H' |\mu\rangle = E_\mu + \langle \mu| V|\mu\rangle + \sum_{\nu \neq \mu} \frac{|\langle \mu | V|\nu\rangle|^2}{E_\mu - E_\nu},
\end{equation}
which is the standard result of non-degenerate perturbation theory [Eq.~\eqref{energy_corrections}]. 

Conversely, if a subspace has dimension larger than 1, but all energies are equal, $E_{\mu n} = E_\mu$, we recover degenerate perturbation theory
\begin{equation}
    \langle \mu n | H' |\mu m\rangle = E_\mu \delta_{n,m} + \langle \mu n| V | \mu m\rangle + \sum_{\nu \neq \mu}\sum_{j\in \nu} \frac{\langle \mu n | V |\nu j \rangle\langle \nu j | V | \mu m\rangle}{E_\mu - E_\nu}.
\end{equation}
Many quantum mechanics textbooks will stop at first order in this degenerate case; the term $\langle \mu n | V | \mu m \rangle$ then tells us how the perturbation breaks (lifts) the degeneracy.
We can also write the degenerate case more compactly in terms of the projectors $P_\mu$
\begin{equation}\label{SW_Hprime_degenerate}
\begin{aligned}
    S &= \sum_{\nu\neq \mu}  \frac{P_\mu V P_\nu}{E_\mu-E_\nu}
    \\[0.2cm]
    H' &= H_0 +  \sum_\mu P_\mu V P_\mu +  \sum_{\mu, \nu \neq \mu} \frac{P_\mu V P_\nu V P_\mu}{E_\mu-E_\nu}
    \\[0.2cm]
    &= H_0 +  \sum_\mu P_\mu V P_\mu +  \sum_{\mu} P_\mu V \frac{1}{E_\mu-H_0}V P_\mu.
\end{aligned}
\end{equation}
This is not possible for Eq.~\eqref{Hprime_matrix_elements} because the energies $E_{\mu n}$ are not all equal for the same $\mu$. 

\subsection{Qutrit example}\label{sec:qutrit_example}

As a simple example, consider a qutrit with 
\begin{equation}\label{AC_example_V}
    H_0 = \begin{pmatrix}
        E_0 & 0 & 0 \\ 
        0 & E_1 & 0 \\
        0 & 0 & E_2
    \end{pmatrix},
    \qquad
    V = \begin{pmatrix}
        0 & v_{01} & 0 \\ 
        v_{01}^* & 0 & v_{12} \\ 
        0 & v_{12}^* & 0
    \end{pmatrix}.
\end{equation}
Suppose levels 0 and 1 are close to each other in energy, but far  from level 2. 
We can therefore identify two subspaces, defined by projectors 
\begin{equation}
    P_{01} = \begin{pmatrix}
        1 & 0 & 0 \\ 
        0 & 1 & 0 \\
        0 & 0 & 0
    \end{pmatrix},
    \qquad 
    P_2 = \begin{pmatrix}
        0 & 0 & 0 \\ 
        0 & 0 & 0 \\
        0 & 0 & 1
    \end{pmatrix}.
\end{equation}
Because of this choice of bundling, 
\begin{equation}
    V_d = \begin{pmatrix}
        0 & v_{01} & 0 \\ 
        v_{01}^* & 0 & 0 \\ 
        0 & 0 & 0
    \end{pmatrix},
    \qquad 
    V_o = \begin{pmatrix}
        0 & 0 & 0 \\ 
        0 & 0 & v_{12} \\ 
        0 & v_{12}^* & 0
    \end{pmatrix}.
\end{equation}
The solution of Eq.~\eqref{S_equation} is 
\begin{equation}\label{AC_stark_S}
        S         
        = \begin{pmatrix}
            0 & 0 & 0 \\
            0 & 0 & -\frac{v_{12}}{\omega_{21}} \\
            0 & \frac{v_{12}^*}{\omega_{21}} & 0 
        \end{pmatrix},
\end{equation}
which, plugging into Eq.~\eqref{Hprime_final_d}, yields
\begin{equation}\label{AC_stark_Hprime}
    H' = \begin{pmatrix}
        E_0 & v_{01} & 0 \\
        v_{01}^* & E_1 - \frac{|v_{12}|^2}{\omega_{21}} & 0 \\
        0 & 0 & E_2+\frac{|v_{12}|^2}{\omega_{21}} 
    \end{pmatrix},
\end{equation}
where $\omega_{ij} := E_j - E_i$. 
In this transformed frame $H'$ is block diagonal in the two subspaces $P_{01}$ and $P_2$.
The coupling $v_{12}$ enters as an effective ``repulsion'', pushing level 1 and 2 further apart. 

\section{Eigenoperator decomposition}\label{sec:eigenoperator_decomposition}

The pathways to apply the SW transformation are now clear. 
Eq.~\eqref{Hprime_matrix_elements} requires clumsy sums over eigenvalues, which become prohibitive for large dimensional problems, such as many-body systems. 
Conversely, the operator approach relies on being able to solve Eq.~\eqref{S_equation}, which involves some reverse engineering: i.e., guessing what is the operator $S$ which, when commuted with $H_0$  lead to $V_o$. 
In this section we will introduce the concept of eigenoperators, which allows one to find a simple and intuitive solution to Eqs.~\eqref{S_equation} and~\eqref{Hprime_final_not_d}. 

An operator $c$ is said to be an eigenoperator of $H_0$ with frequency $\omega$ if 
\begin{equation}\label{eigenops_def}
    [H_0,c] = -\omega c.
\end{equation}
An eigenoperator with positive frequency $\omega>0$ acts as a lowering operator for $H_0$, taking an eigenvector with energy $E$ to another with energy $E-\omega$. 
From Eq.~\eqref{eigenops_def} it follows that 
\begin{equation}\label{eigenops_def_dagger}
    [H_0,c^\dagger] = \omega c^\dagger,    
\end{equation}
so $c^\dagger$ is also an eigenoperator, but with frequency $-\omega$. 
Eigenoperators therefore have to be non-Hermitian,  unless $\omega = 0$. 
The converse is also true: a Hermitian operator can only be an eigenoperator if it has zero frequency (i.e., commutes with $H_0$).
Combining Eqs.~\eqref{eigenops_def} and~\eqref{eigenops_def_dagger} we find that $[H_0,c^\dagger c] =0$, but note that, in general, nothing can be said about $[c,c^\dagger]$.
A useful property of eigenoperators is that
\begin{equation}\label{eigenops_c_time_dependence}
    e^{i H_0 t} c e^{-i H_0 t} = e^{-i \omega t} c,
\end{equation}
which follows directly from Eq.~\eqref{eigenops_def} and the Baker-Campbell-Hausdorff formula.

Familiar examples of eigenoperators include 
\begin{equation}
    \begin{aligned}
        H_0 &= 
        \frac{\omega}{2}\sigma_z ,\qquad  c = g\sigma_-
        \\[0.2cm]
        H_0 &= \omega a^\dagger a, \qquad  c = ga,
    \end{aligned}
\end{equation}
where $g$ is an arbitrary constant, 
$\sigma_i$ are Pauli matrices and $a$ is a bosonic or fermionic operator.
As another example, consider a qutrit with 
\begin{equation}\label{qutrit_example}
    H_0 = \begin{pmatrix}
        E_0 & 0 & 0 \\
        0 & E_1 & 0 \\
        0 & 0 & E_2
    \end{pmatrix} = \sum_j E_j \sigma_j.
\end{equation}
Here and henceforth, when dealing with qudits, we will use the notation 
\begin{equation}
    \sigma_{ij} = |i\rangle\langle j|,
    \qquad 
    \sigma_j = \sigma_{jj} = |j\rangle\langle j| .
\end{equation}
One can readily verify that, e.g.,
\begin{equation}\label{qutrit_c}
    c=\begin{pmatrix}
     0 & c_{01} & 0 \\ 
     0 & 0 & 0 \\ 
     0 & 0 & 0
    \end{pmatrix} = c_{01} \sigma_{01},
\end{equation}
is an eigenoperator of $H_0$ with frequency $\omega = E_1 - E_0$, for any value of $c_{01}$.
More generally,  $\alpha \sigma_{ij}$ is an eigenoperator with frequency $E_j - E_i$, for any $\alpha$. 
A combination such as $c_{01} \sigma_{01} + c_{12} \sigma_{12}$ will, in general,  \emph{not} be an eigenoperator, unless the transition frequencies are degenerate ($E_2 - E_1= E_1 - E_0$).
It is important to emphasize that what matters is the degeneracy of the transition frequencies, not the degeneracies of the energies themselves. 

The above results may seem somewhat trivial. 
But their usefulness lies in the fact that any operator $V$ can be decomposed as a sum of eigenoperators:
\begin{equation}\label{V_eigenoperator_decomposition}
    V = \sum_{\omega} V_\omega = \sum_{\omega}V_\omega^\dagger,
\end{equation}
where the sum is over all possible transition frequencies $\omega = E_{\nu m} - E_{\mu n}$ of $H_0$,
and each $V_\omega$ satisfies 
\begin{equation}\label{V_omega_H0_commutation}
    [H_0,V_\omega] = -\omega V_\omega,
    \qquad 
    V_\omega^\dagger = V_{-\omega}^{}.
\end{equation}
The two ways of writing Eq.~\eqref{V_eigenoperator_decomposition} are  equivalent since the sum contains positive, negative and zero frequencies. 

The explicit formula for $V_\omega$ is 
\begin{equation}\label{eigenoperator_explicit_formula}
    V_\omega = \sum_{\mu,n} \sum_{\nu,j} |\mu n\rangle\langle \mu n| V |\nu j\rangle\langle \nu j| \delta_{\omega, E_{\nu j} - E_{\mu n}}.
\end{equation}
where the Kronecker delta selects the matrix elements of $V$ for which the energy difference is exactly $\omega$. 
In practice, however, this formula is seldom useful, and a much more convenient way of finding $V_\omega$ will be explained below. 


The relation $V_\omega^\dagger = V_{-\omega}^{}$ is only true when $V$ is Hermitian. 
One can also do the eigenoperator decomposition for non-Hermitian operators. However, in this case $[V_\omega]^\dagger \neq [V^\dagger]_\omega$, which can be  confusing. 
Henceforth, when this is the case, we will always take $V_\omega^\dagger$ to mean $[V_\omega]^\dagger$.

\subsection{Procedure for finding eigenoperators}\label{sec:procedure_eigenops}

The only general formula for finding eigenoperators is Eq.~\eqref{eigenoperator_explicit_formula}. 
It turns out, however, that there are many cases where one can ``guess'' what the eigenoperators are using the following trick. 
As discussed in Eq.~\eqref{eigenops_c_time_dependence}, $V_\omega$ will evolve under $H_0$ as $e^{-i \omega t}$. Hence, it follows from Eq.~\eqref{V_eigenoperator_decomposition} that 
\begin{equation}\label{V_time_evolution}
    e^{i H_0 t} V e^{-i H_0 t} = \sum_\omega e^{-i \omega t} V_\omega.
\end{equation}
Therefore, if we can compute $e^{i H_0 t} V e^{-i H_0 t}$, we simply need to identify all distinct exponentials $e^{-i\omega t}$, and the operator multiplying each one will be $V_\omega$.

For example, suppose 
\begin{equation}\label{basic_bosonic_model_H0_V}
 H_0 = \omega_r a^\dagger a, 
 \qquad 
 V = g(a+a^\dagger),
\end{equation}
where $a$ is a bosonic mode, and $\omega_r$ its frequency. From the BCH formula we know that 
\begin{equation}
    e^{i \omega_r t a^\dagger a} a e^{-i \omega_r t a^\dagger a} = e^{-i \omega_r t} a.
\end{equation}
Thus
\begin{equation}
    e^{i H_0 t} V e^{-i H_0 t} = e^{-i \omega_r t} g a + e^{i \omega_r t} g a^\dagger. 
\end{equation}
The eigenoperator decomposition~\eqref{V_eigenoperator_decomposition} is therefore 
\begin{equation}
 V = g(a+a^\dagger) = V_{\omega_r} + V_{-\omega_r}, 
 \qquad V_{\omega_r} = g a
\end{equation}
and $V_{-\omega_r}^{} = V_{\omega_r}^\dagger$ (notice how we only need to list the positive frequency operators). 

As a second example, suppose we continue to have $H_0 = \omega_r a^\dagger a$ but change the perturbation to 
\begin{equation}
    \begin{aligned}
        V &= g(a+a^\dagger)^2 
        \\[0.2cm]
        &= g(a^2 + a^{\dagger 2} + 2 a^\dagger a + 1)
        \\[0.2cm]
        &= V_{2\omega_r} + V_{2\omega_r}^\dagger + V_0,
    \end{aligned}
\end{equation}
where $V_{2\omega_r} = ga^2$ and $V_0 = g(2a^\dagger a + 1)$. 
Since we know that each $a$ picks up $e^{-i \omega_rt}$ and each $a^\dagger$ picks up $e^{i\omega_r t}$, the eigenoperators could be read from the second line without the need to do any actual calculations.

This approach is very modular and allows us to construct eigenoperators even for complicated perturbations.
All we need to know is how each individual element of $V$ is decomposed, and then we can just combine things together. 
For example, in Ref.~\cite{mercurio2024} the authors recently studied a 3-boson model with 
\begin{equation}\label{Nori_LoFranco_H0_V}
    \begin{aligned}
        H_0 &= \omega_a a^\dagger a + \omega_b b^\dagger b + \omega_c c^\dagger c, 
        \\[0.2cm]
        V &= - \frac{g}{2}\Big[(a+a^\dagger)^2 -\Omega^2 (c+c^\dagger)^2\Big](b+b^\dagger).
    \end{aligned}
\end{equation}
Because each annihilation operator picks up a phase $e^{-i \omega_i t}$, expanding out all the terms in $V$ immediately yields: 
\begin{equation}
\begin{aligned}
    V_{\omega_b-2\omega_a} &= -\frac{g}{2}a^{\dagger 2} b,
    \qquad 
    V_{\omega_b - 2 \omega_c} = \frac{g\Omega^2}{2} c^{\dagger 2} b
    \\[0.2cm]
    V_{\omega_b+2\omega_a} &= -\frac{g}{2} a^2 b,
    \qquad ~
    V_{\omega_b+2\omega_c} = \frac{g\Omega^2}{2}c^2 b
    \\[0.2cm]
    V_{\omega_b} &= -\frac{g}{2}\Big[(2a^\dagger a+1) -\Omega^2(2c^\dagger c+1)\Big] b,
\end{aligned}
\end{equation}
together with their adjoints. 
Table~\ref{tab:eigenoperator_examples} lists various other examples.

\begin{table*}[!t]
    \centering
    \caption{Examples of how to guess the eigenoperator decomposition [Eq.~\eqref{V_eigenoperator_decomposition}]. Here $a,b,c$ are  bosonic operators, $\sigma_\alpha$ are Pauli matrices and $c_{1/2}$ are fermionic operators. Only half of the eigenoperators are listed. The remainder correspond to the adjoints $V_\omega^\dagger$. \\}
    \begin{tabular}{c|c|c|c}    
         Label & $H_0$ & $V$ & Eigenoperator decomposition \\[0.2cm] \hline
         &&&\\[-0.1cm]
       1 boson; linear $V$ &  $H_0=\omega_r a^\dagger a$ 
         & 
         $V = g(a+a^\dagger)~~~~~~~~~~~~~$ & $V_{\omega_r} = g a~~~~~~~~~~$ 
         \\[0.3cm]\hline &&&\\[-0.1cm]
        1 boson  & $H_0=\omega_r a^\dagger a$ 
         & 
         $V = g(a+a^\dagger)^2 ~~~~~~~~~~~$ & $~~~~~~~V_0 = g(2a^\dagger a + 1)$ 
         \\[0.2cm]
      quadratic $V$ &  & $~~~~~~~~~~~= g(2a^\dagger a + 1 + a^2 + a^{\dagger 2})$ & $V_{2\omega_r} = g a^2~~~~~~~~~~$
         \\[0.3cm]\hline &&&\\[-0.1cm]
       1 boson &  $H_0=\omega_r a^\dagger a$ 
         & 
         $V = g(a+a^\dagger)^4 ~~~~~~~~~~~$ & $~~~~~~~~~~~~~~~~~V_0 = g(3 + 12 a^\dagger a+ 6 a^\dagger a^\dagger a a)$ 
         \\[0.2cm]
        quartic $V$ & & & $V_{2\omega_r} = g (6 a^2 + 4 a^{\dagger} a^3)$ 
         \\[0.2cm]
         & & & $V_{4\omega_r} = g a^4$~~~~~~~~~~~~~~~~~~
         \\[0.3cm]\hline &&&\\[-0.1cm]
        2 bosons &  $H_0=\omega_a a^\dagger a + \omega_b b^\dagger b$ 
         & 
         $V = g(a+a^\dagger)(b+b^\dagger) ~~~~~~~~~~~$ & $V_{\omega_a-\omega_b} = g a b^\dagger~~~~~~~~~~$ 
         \\[0.2cm]
       quadratic $V$ &  & & $V_{\omega_a+\omega_b} = g a b$~~~~~~~~~~~~
         \\[0.2cm]\hline &&& \\[-0.1cm]
        2 bosons & $H_0=\omega_a a^\dagger a + \omega_b b^\dagger b$ 
         & 
         $V = g(a+a^\dagger)^2(b+b^\dagger) ~~~~~~~~~~~$ & $V_{\omega_b-2\omega_a} = g a^{\dagger 2}b~~~~~~~~~~$ 
         \\[0.2cm]
       Cubic $V$ &  & & $V_{\omega_b+2\omega_a} = g a^2 b$~~~~~~~~~~~~~~
         \\[0.2cm]
       &  & &~~~~ $V_{\omega_b} = g (2a^\dagger a+1) b$
         \\[0.2cm]\hline\hline &&& \\[-0.1cm]             
       Jaynes-Cummings &  $H_0 = \frac{\omega_q}{2}\sigma_z + \omega_r a^\dagger a$& $V = g (\sigma_+ a + a^\dagger \sigma_-)$
         & $V_{\omega_q-\omega_r} = g \sigma_- a^\dagger$~~~~~~~~~~~~
         \\[0.2cm]\hline &&& \\[-0.1cm]
        Rabi &  $H_0 = \frac{\omega_q}{2}\sigma_z + \omega_r a^\dagger a$
         & $V = g \sigma_x(a+a^\dagger)~~~~~~~~~~~~~~~~~~~~~~~~~~~~~~$
         & $V_{\omega_q-\omega_r} = g \sigma_- a^\dagger$~~~~~~~~~~~~
         \\[0.2cm] 
        & & $~~= g(\sigma_+ a + \sigma_- a^\dagger + \sigma_+ a^\dagger + \sigma_- a)$ & $V_{\omega_q+\omega_r} = g \sigma_- a~~$ ~~~~~~~~~~~~
         \\[0.2cm]\hline 
         \hline &&& \\[-0.3cm]
       Interacting &  $H_0 = \sum_{i=1,2} \epsilon_i \hat{n}_i + U_{12} \hat{n}_1 \hat{n}_2$
        & $V = c_1$
         & $V_{\epsilon_1-U} = c_1 \hat{n}_2 $
         \\[0.2cm]
       fermions & & & ~~~~$V_{\epsilon_1} = c_1 \hat{h}_2$
         \\[0.2cm]\hline &&& ~~~~~~~~~~~~~~\\[-0.1cm]
    \end{tabular}
    \label{tab:eigenoperator_examples}
\end{table*}

\subsection{Eigenoperators for interacting fermions}\label{sec:eigen_ops_fermi_Hubbard}

The method described above works very often, but not always. 
In the original Schrieffer-Wolff paper~\cite{schrieffer1966} an alternative method was implicitly used, which applies to interacting fermionic systems. 
This method also relates with the so-called Hubbard operators (see~\cite{fazekas} for a details).
Consider a system with two fermionic modes $c_1$ and $c_2$, and Hamiltonian 
\begin{equation}\label{Hubbard_2}
    H_0 = \sum_{i=1,2} \epsilon_i \hat{n}_i + U \hat{n}_1 \hat{n}_2,
\end{equation}
where $\hat{n}_i = c_i^\dagger c_i^{}$. 
This could be, e.g. a single fermionic site with $c_1 = c_\uparrow$ and $c_2 = c_\downarrow$.
The first term describes the on-site potentials and the second describes the 2-body Fermi-Hubbard interactions. 
The operators $c_i$ are eigenoperators of the first term, $\sum_i \epsilon_i \hat{n}_i$, because  $[\hat{n}_i,c_i] = -c_i$.
However, they are not eigenoperators of $U \hat{n}_1 \hat{n}_2$.
To decompose $c_i$ into eigenoperators we introduce the hole operator, $\hat{h}_i = 1- \hat{n}_i$, and write 
\begin{equation}\label{fermion_decomposition_ci}
    c_1 = c_1 \hat{n}_2 + c_1 \hat{h}_2.
\end{equation}
Then one can readily verify that 
\begin{equation}\label{38198398189729817}
    \begin{aligned}
        [H_0, c_1 \hat{n}_2] &= -(\epsilon_1 + U) c_1 \hat{n}_2,
        \\[0.2cm]
        [H_0,c_1\hat{h}_2] &= -\epsilon_1 c_1 \hat{h}_2,
    \end{aligned}
\end{equation}
so that $c_1 \hat{n}_2$ and $c_1 \hat{h}_2$ \emph{are} eigenoperators of $H_0$. 
The decomposition for $c_2$ would be similar, but with $\hat{n}_1$ and $\hat{h}_1$. 

It is straightforward to generalize this to an arbitrary number $N$ of fermions with Hamiltonian
\begin{equation}\label{Fermi_Hubbard_H0}
    H_0 = \sum_i \epsilon_i \hat{n}_i + \sum_{j>i} U_{ij} \hat{n}_i \hat{n}_j,
\end{equation}
We will simply have:
\begin{equation}
    c_i = c_i \prod_{j\neq i} (\hat{n}_j + \hat{h}_j).
\end{equation}
For example, if $N = 3$
\begin{equation}
    c_1 = c_1 \hat{n}_2\hat{n}_3 + c_1 \hat{n}_2\hat{h}_3 + c_1 \hat{h}_2\hat{n}_3 + c_1 \hat{h}_2 \hat{h}_3
\end{equation}
The complexity, of course, quickly increases with the number of pairs of modes that interact.

Coming back to the two-fermion problem~\eqref{Hubbard_2}, the Hilbert space basis for two fermions only can be written as 
\begin{equation}\label{two_fermion_basis}
    |0\rangle, 
\quad 
|1\rangle = c_1^\dagger |0\rangle,
\quad 
|2\rangle = c_2^\dagger |0\rangle,
\quad 
|d\rangle = c_1^\dagger c_2^\dagger |0\rangle.
\end{equation}
Recall also that the vacuum can be represented as 
\begin{equation}
    |0\rangle\langle 0| = \hat{h}_1 \hat{h}_2. 
\end{equation}
We now introduce the Hubbard operators 
\begin{equation}\label{hubbard_op_def}
    X_{ij} = |i\rangle\langle j|, \qquad i,j = 0,1,2,d.
\end{equation}
The diagonal Hubbard operators are 
\begin{equation}
    X_{00} = \hat{h}_1\hat{h}_2, 
    \quad 
    X_{11} = \hat{n}_1\hat{h}_2,
    \quad 
    X_{22} = \hat{n}_2 \hat{h}_1,
    \quad 
    X_{dd} = \hat{n}_1\hat{n}_2.
\end{equation}
and the off-diagonal ones are 
\begin{align}\label{hubbard_ops_off_diagonal}
X_{10} &= c_1^\dagger h_2,
\qquad 
X_{20} = c_2^\dagger h_1,
\qquad 
X_{d0} = c_1^\dagger c_2^\dagger,
\\[0.2cm]
X_{21} &= c_2^\dagger c_1^{},
\qquad 
X_{d1} = -n_1 c_2^\dagger ,
\qquad
X_{d2} = n_2 c_1^\dagger ,
\end{align}
as well as $X_{ji}^{} = X_{ij}^\dagger$.
The Hubbard operators are all eigenoperators of $H_0$ in Eq.~\eqref{Hubbard_2}: 
\begin{align}
[H,X_{ii}] &= 0
\\
[H,X_{i0}] &= \epsilon_i X_{i0},\qquad i = 1,2
\\
[H,X_{d0}] &= (\epsilon_1+\epsilon_2 + U) X_{d0}
\\
[H,X_{21}]&= (\epsilon_2-\epsilon_1) X_{21}
\\
[H,X_{d1}] &= (\epsilon_2+U) X_{d1}
\\
[H,X_{d2}] &= (\epsilon_1+U) X_{d2}
\end{align}
Eq.~\eqref{fermion_decomposition_ci} is a particular case of this:
\begin{equation}
    c_1^\dagger = X_{10} + X_{d2},
    \qquad 
    c_2^\dagger = X_{20} - X_{d1},
\end{equation}
which can be read directly from Eq.~\eqref{hubbard_ops_off_diagonal}.
The minus sign comes from the choice of ordering in the state $|d\rangle$ in Eq.~\eqref{two_fermion_basis}.
Written in this way, this makes it clearer that Eq.~\eqref{fermion_decomposition_ci} is, in fact, a decomposition of $c_1^\dagger$ into a creation of fermion 1 when 2 is empty, plus a creation of fermion 1 when 2 is occupied. 

\section{Schrieffer-Wolff transformation in terms of eigenoperators}\label{sec:main}

\begin{figure}
    \centering
    \includegraphics[width=0.4\textwidth]{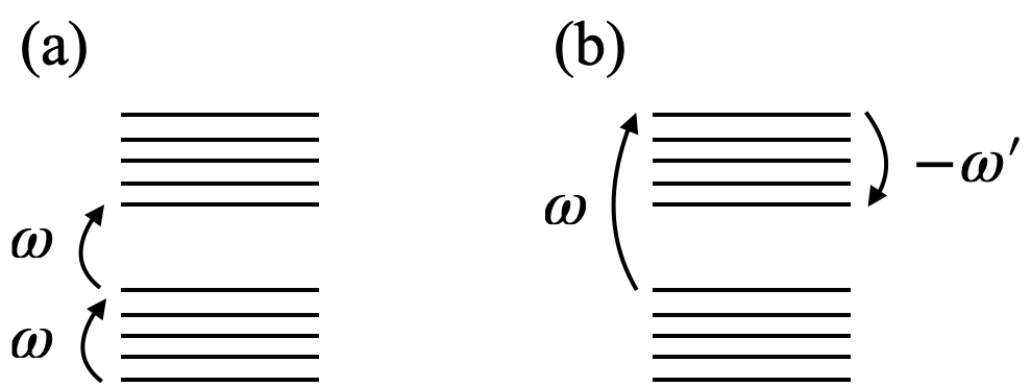}
    \caption{Some caveats that might appear when using the eigenoperator decomposition to write the Schrieffer-Wolff transformation. (a) An eigenoperator $V_\omega$ can sometimes generate both block diagonal and block off-diagonal transitions. In this case one must split it as $V_\omega = V_{d,\omega} + V_{o,\omega}$, and include the former in $V_d$ and the latter in $V_o$. 
    (b) In Eq.~\eqref{Hprime_eigen_final_d}, to obtain a block diagonal Hamiltonian we only keep terms which combine $V_{o,\omega}$ with $V_{o,\omega'}^\dagger$, and therefore generate a net transition with frequency $\omega-\omega'$ (for $\omega,\omega'>0$). In some cases $\omega-\omega'$ might not land you back to the original subspace, and is therefore not block diagonal. These terms must be neglected, which is denoted in Eq.~\eqref{Hprime_eigen_final_d} by the prime in the sum. 
    }
    \label{fig:caveats}
\end{figure}

The usefulness of the eigenoperator decomposition in the Schrieffer-Wolff transformation is that it will allow us to trivially solve for Eq.~\eqref{S_equation}.
To start, we expand $V$ in terms of eigenoperators, exactly as in Eq.~\eqref{V_eigenoperator_decomposition}. 
To apply the SW transformation we need to split $V = V_d + V_o$; that is, choose which $V_\omega$ should belong to $V_d$ and which should belong to $V_o$.
Recall that $\omega$ runs over the set of allowed transition frequencies $E_{\nu m} - E_{\mu n}$ of $H_0$. 
So terms with $\omega$ belonging to the same subspace (including $\omega=0$) should go to $V_d$, and those that go between two subspaces to $V_o$. 
In some special cases it might be necessary to further split  $V_\omega$ in two parts. For example, in the case illustrated in Fig.~\ref{fig:caveats}(a), a given $V_\omega$ will generate transitions within the same subspace, as well as between two subspaces. If this is the case we need to further split $V_\omega = V_{d,\omega} + V_{o,\omega}$ and place one bit in $V_d$ and the other in $V_o$.

From this we emerge with an eigenoperator decomposition of $V_o$, which has the form
\begin{equation}\label{Vo_eigen_op}
    V_o = \sum_{\omega>0} (V_{o,\omega}^{} + V_{o,\omega}^\dagger).
\end{equation}
Once this decomposition is found, the application of the SW method is actually done: 
because of Eq.~\eqref{V_omega_H0_commutation}, 
the solution of Eq.~\eqref{S_equation} is
\begin{equation}\label{S_sol}
    S = - \sum_{\omega>0}\tfrac{1}{\omega}(V_{o,\omega}^{}-V_{o,\omega}^\dagger).
\end{equation}
Next we plug this in Eq.~\eqref{Hprime_final_not_d}. Note that 
\begin{equation}\label{038719871398719871}
    [S,V_o]= - \sum_{\omega,\omega' >0} \frac{1}{\omega}[V_{o,\omega}^{}-V_{o,\omega}^\dagger, V_{o,\omega'}^{} + V_{o,\omega'}^\dagger].    
\end{equation}
To get to Eq.~\eqref{Hprime_final_d} the final step is to take the block diagonal part $[S,V_o]_d$.
But notice that the product of two eigenoperators is also an eigenoperator
\begin{equation}
    [H_0, V_{o,\omega} V_{o,\omega'}] = -(\omega+\omega') V_{o,\omega^{}} V_{o,\omega'}.
\end{equation}
The terms $[V_{o,\omega^{}},V_{o,\omega'}]$ and $[V_{o,\omega^{}}^\dagger, V_{o,\omega'}^\dagger]$ in Eq.~\eqref{038719871398719871} therefore generate double transitions with $\omega+\omega'$, which are block off-diagonal and do not contribute to $[S,V_o]_d$. 
Conversely, the terms $[V_{o,\omega}^{},V_{o,\omega'}^\dagger]$ and $[V_{o,\omega}^\dagger, V_{o,\omega'}^{}]$ generate transitions with frequency $\omega-\omega'$. 
These are also not guaranteed to land you back on the same subspace (c.f.~Fig.~\ref{fig:caveats}(b)). 
We therefore must also drop all terms where $\omega-\omega'$ is block off-diagonal. 
With this caveat, the Schrieffer-Wolff Hamiltonian finally becomes
\begin{equation}\label{Hprime_eigen_final_d}
    H' = H_0 + V_d - \tfrac{1}{2}\sideset{}{'}\sum_{\omega,\omega'>0}^{} \left( \tfrac{1}{\omega}+\tfrac{1}{\omega'}\right) [V_{o,\omega}^{}, V_{o,\omega'}^\dagger],
\end{equation}
where the prime in the summation means one should exclude any pairs for which $\omega-\omega'$ is not block diagonal. 
Eq.~\eqref{Hprime_eigen_final_d} is the main result of the paper. 
It provides a systematic and intuitive way of building SW Hamiltonians. 
And it is entirely equivalent to Eq.~\eqref{Hprime_matrix_elements}, but recast in terms of eigenoperators. 
Notice how the corrections involve transitions that take the system away of a subspace by $\omega$, and then back by $\omega'$.
These are precisely the virtual transitions. 
Below, whenever convenient, we will also use the more compact notation 
\begin{equation}\label{f_coeff}
    f_{\omega,\omega'} = \tfrac{1}{2}\left( \tfrac{1}{\omega}+\tfrac{1}{\omega'}\right),
\end{equation}
for the coefficient in the last term of~\eqref{Hprime_eigen_final_d}.

\section{Examples}
\subsection{A single bosonic mode}\label{sec:single_boson}

Consider the bosonic model in Eq.~\eqref{basic_bosonic_model_H0_V}. 
As we saw in Table~\ref{tab:eigenoperator_examples},  $V_{\omega_r} = g a$. 
There is only one term in the eigenoperator decomposition, and it should belong to $V_o$.  Eq.~\eqref{S_sol} then reads 
\begin{equation}
    S = \frac{g}{\omega_r}(a^\dagger-a),
\end{equation}
while Eq.~\eqref{Hprime_eigen_final_d} yields 
\begin{equation}
    H' = \omega_r a^\dagger a - \frac{g^2}{\omega_0} [a,a^\dagger]= \omega_r a^\dagger a- \frac{g^2}{\omega}.
\end{equation}
In this case the result turns out to be exact because the transformation unitary $e^{S}$ of Eq.~\eqref{Hprime_def} is a displacement operator, which actually diagonalizes $H_0 + V$. 

Next suppose we change the perturbation to $V = g(a+a^\dagger)^2$. 
Then $V_{2\omega_r} = ga^2$ and $V_0 = g(2a^\dagger a + 1)$. 
The former belongs to $V_o$, while the latter belongs to $V_d$, since it generates no transitions. 
We therefore get from Eq.~\eqref{S_sol} 
\begin{equation}
S = \tfrac{g}{2\omega_r}(a^2 - a^{\dagger 2}),    
\end{equation}
which is a single mode squeezing. 
The SW Hamiltonian [Eq.~\eqref{Hprime_eigen_final_d}] is
\begin{equation}
    \begin{aligned}
        H' &= \omega_r a^\dagger a + g (2a^\dagger a + 1) - \frac{g^2}{2\omega_r} [a^2,a^{\dagger2}] 
    \\[0.2cm]
    &= \Big(\omega_r + 2 g - \frac{2g^2}{\omega_r}\Big)a^\dagger a + g - \frac{g^2}{\omega_r}.
    \end{aligned}
\end{equation}
The result is now only approximate, and will hold for $g \ll \omega_r$.

As a third (more interesting) example, consider the Duffing oscillator  
\begin{equation}
    H_0 = \omega_r a^\dagger a,\qquad 
    V = g(a+a^\dagger)^4.
\end{equation}
The eigenoperators are obtained by the same method discussed in Sec.~\ref{sec:procedure_eigenops}, and are shown in Table~\ref{tab:eigenoperator_examples}. 
The two operators $V_{2\omega_r}$ and $V_{4\omega_r}$ should both belong to $V_o$. 
But in Eq.~\eqref{Hprime_eigen_final_d} we should discard terms such as $[V_{2\omega_r}^{}, V_{4\omega_r}^\dagger]$, since these will generate transitions that are not block diagonal (this is precisely the prime in the sum in Eq.~\eqref{Hprime_eigen_final_d}). 
The SW Hamiltonian therefore becomes 
\begin{equation}
    H' = \omega_r a^\dagger a + V_d - \frac{1}{2\omega_r} [V_{2\omega_r}^{}, V_{2\omega_r}^\dagger]
    - \frac{1}{4\omega_r} [V_{4\omega_r}^{}, V_{4\omega_r}^\dagger].
\end{equation}
The leading order correction in $g$ is actually 
\begin{equation}
    V_d \equiv V_{\omega=0} =g(3 + 12 a^\dagger a+ 6 a^{\dagger2}a^2).
\end{equation}
This therefore gives a correction to the system energy, plus the typical Kerr non-linearity $a^{\dagger2} a^2$ that is obtained by doing a rotating wave approximation. 
To obtain an explicit formula for the second order term, we need to carry out the cumbersome commutation relations 
\begin{equation}
    \begin{aligned}
        [V_{2\omega_r}^{}, V_{2\omega_r}^\dagger] &= 8g^2 \Big( 9 + 66 a^\dagger a + 72 a^{\dagger 2} a^2 + 16 a^{\dagger 3} a^3\Big),
        \\[0.2cm]        [V_{4\omega_r}^{},V_{4\omega_r}^\dagger] &= 8 g^2 \Big(3 + 12 a^\dagger a + 9 a^{\dagger 2} a^2 + 2 a^{\dagger 3} a^3
        \Big).
    \end{aligned}
\end{equation}
We then get
\begin{equation}
    \begin{aligned}
        H' =& \omega_r a^\dagger a + g (3 + 12 a^\dagger a + 6 a^{\dagger 2} a^2) 
        \\[0.2cm]
        &- \frac{2g^2}{\omega_r}\big(21 + 144 a^\dagger a + 153 a^{\dagger 2} a^2 + 34 a^{\dagger 3} a^3\big).
    \end{aligned}
\end{equation}
There are, therefore, further corrections to both the quadratic and the Kerr term, plus the appearance of a 6th order contribution.
The resulting Hamiltonian is once again already diagonal, and one can verify that the resulting energies coincide with what one would obtain from either standard perturbation theory, or from numerical diagonalization. 


\subsection{Two bosonic modes}
\label{sec:two_bosons}

Consider two bosonic modes $a$ and $b$, with
\begin{equation}\label{example_two_boson_model}
    \begin{aligned}
        H_0 &= \omega_a a^\dagger a + \omega_b b^\dagger b
        \\[0.2cm]
        V &= g(a+a^\dagger)(b+b^\dagger)
    \end{aligned}
\end{equation}
This problem is interesting because there are 3 competing energy scales, $\omega_a, \omega_b$ and $g$. 
The corresponding eigenoperators are (c.f.~Table~\ref{tab:eigenoperator_examples})
\begin{equation}
    \begin{aligned}
        V_{\omega_a-\omega_b} &= g a b^\dagger,
        \\[0.2cm]
        V_{\omega_a+\omega_b} &= g a b.
    \end{aligned}
\end{equation}
To apply the SW transformation the first step is to choose which operators should belong to $V_d$ and which should belong to $V_o$.
Clearly $V_{\omega_a+\omega_b}$ will belong to $V_o$. 
But the choice of where to place $V_{\omega_a-\omega_b}$ depends on how close $\omega_a$ is from $\omega_b$. 

\hl{Quasi-degenerate modes ($\omega_a\simeq \omega_b$):} in this case $V_{\omega_a-\omega_b}$ belongs to $V_d$. 
This amounts to taking the subspaces as two-dimensional blocks spanned by Fock states with equal occupation, $|n,n\rangle$. 
The only element of $V_o$ is thus $V_{\omega_a+\omega_b} = g ab$.
Eq.~\eqref{S_sol} gives us 
\begin{equation}
    S = - \frac{g}{\omega_a+\omega_b} (ab -a^\dagger b^\dagger),
\end{equation}
which is a 2-mode squeezing. 
The SW Hamiltonian~\eqref{Hprime_eigen_final_d} yields \begin{equation}
\begin{aligned}
    H' =& H_0 + g(a^\dagger b + b^\dagger a)- \frac{g^2}{\omega_a+\omega_b} [ab^\dagger,a^\dagger b]
    \\[0.2cm]
    =& H_0 + g(a^\dagger b + b^\dagger a) - \frac{g^2}{\omega_a+\omega_b}(b^\dagger b - a^\dagger a).
\end{aligned}    
\end{equation}
This is not yet diagonal because of the terms linear in $g$. The effect of the $\omega_a+\omega_b$ transitions is seen to be just a renormalization of the two modes by $\pm g^2/(\omega_a+\omega_b)$. 

\hl{Far off-resonance modes ($\omega_a\gg \omega_b$):} 
This is called the dispersive regime and will be studied more systematically in Sec.~\ref{sec:dispersive}. 
Because they are far from resonance, mode $b$ is unable to generate transitions in $a$. 
Each subspace therefore corresponds to a fixed Fock state of $a$. 
That is, in the language of Eq.~\eqref{H0_eigen} 
$|\mu,n\rangle \to |n_a,n_b\rangle$ [i.e., for each Fock state of $a$, there are many Fock states of $b$.]
As a consequence, both $V_{\omega_a+\omega_b}$ and $V_{\omega_a-\omega_b}$ will belong to $V_o$.
The basis transformation matrix~\eqref{S_sol} will then be 
\begin{equation}
    S = - \frac{g}{\omega_a+\omega_b}(ab - a^\dagger b^\dagger) - \frac{g}{\omega_a-\omega_b}(ab^\dagger - a^\dagger b),
\end{equation}
which is a combination of a two-mode squeezing and a beam splitter. 

To obtain the SW Hamiltonian in Eq.~\eqref{Hprime_eigen_final_d} we need to work out all four combinations shown in Table~\ref{tab:two_bosons}. 
As a result, we find
\begin{equation}\label{example_two_bosons_no_RWA}
\begin{aligned}
    H' =& H_0  - \frac{g^2}{\omega_a-\omega_b}(b^\dagger b - a^\dagger a)
    \\[0.2cm]
    & - \frac{g^2}{\omega_a+\omega_b}(a^\dagger a + b^\dagger b +1)- \frac{g^2\omega_a}{\omega_a^2-\omega_b^2} (b^2 + b^{\dagger 2}).
\end{aligned}    
\end{equation}
We see here two tiers of approximations, which will depend on the interplay between $\omega_b$ and $g$. 
We are already assuming $\omega$ is the largest scale: $\omega_a \gg \omega_b,g$. 
If $g \ll \omega_b$ then the second line will contribute much less and can be neglected. This is tantamount to doing a rotating wave approximation in Eq.~\eqref{example_two_boson_model}. 
Conversely if $g$ is not much smaller than $\omega_b$, we are essentially saying that the transition $2\omega_b$ should still be considered to be block diagonal. 
In this case, therefore, there will be an additional single-mode squeezing to mode $b$ (the last term).

\begin{table}[!t]
    \centering
    \caption{Terms in the sum in Eq.~\eqref{Hprime_eigen_final_d} for $V_{\omega_a+\omega_b} = g ab$ and $V_{\omega_a-\omega_b} = g a b^\dagger$.}
    \begin{tabular}{c|c|c|c|c}
    $\omega$ & $\omega'$ & $\omega - \omega'$ & $\frac{1}{2}\left(\frac{1}{\omega}+\frac{1}{\omega'}\right)$ & $[V_{o,\omega}, V_{o,\omega'}^\dagger]$ \\[0.2cm]\hline
    $\omega_a+\omega_b$ & $\omega_a+\omega_b$ & 0 & $\frac{1}{\omega_a+\omega_b}$ & $g^2(a^\dagger a + b^\dagger b + 1)$ \\[0.2cm]
     $\omega_a+\omega_b$ & $\omega_a-\omega_b$ & $2 \omega_b$ & $\frac{ \omega_a}{\omega_a^2-\omega_b^2}$ & $g^2 b^2$ \\[0.2cm]
     $\omega_a-\omega_b$ & $\omega_a+\omega_b$ & $-2 \omega_b$ & $\frac{ \omega_a}{\omega_a^2-\omega_b^2}$ & $g^2 b^{\dagger2}$ \\[0.2cm]
     $\omega_a-\omega_b$ & $\omega_a-\omega_b$ & 0 & $\frac{1}{\omega_a-\omega_b}$ & $g^2(b^\dagger b -a^\dagger a)$
     \end{tabular}    
    \label{tab:two_bosons}
\end{table}

\section{Dispersive interactions}\label{sec:dispersive}

Suppose we have two systems interacting with a total Hamiltonian 
\begin{equation}
    H = H_A + H_B + V := H_0 + V.
\end{equation}
We write 
\begin{equation}
 H_A = \sum_a E_a |a\rangle\langle a|,
 \qquad 
H_B = \sum_b E_b|b\rangle\langle b|.
\end{equation}
One of the most important uses the SW method is to describe the so-called dispersive regime, in which the energy spacings of $A$ are much larger than those of $B$, 
\begin{equation}
|E_{a'}-E_a| \gg |E_{b'}- E_b|.
\end{equation}
In other words, the two systems are very far off-resonance. 
As a consequence, $B$ is essentially unable to excite $A$, which will therefore remain in whatever initial energy subspace it starts in~\cite{CohenTannoudji1998}.

The energy diagram for $H_0 = H_A + H_B$ is drawn in Fig.~\ref{fig:drawing}(b).
The connection with the original SW scenario of Fig.~\ref{fig:drawing}(a) is now clear: each energy subspaces $\mu$ corresponds to an energy eigenvalue of $A$. 
And the states within that subspace correspond to all possible states of $B$. That is, in the language of Eq.~\eqref{H0_eigen}, we have $|a b\rangle$ instead of $|\mu n\rangle$.
With this change, one can easily adapt Eq.~\eqref{Hprime_matrix_elements} to write down the matrix elements of the effective Hamiltonian. 
However, as before, this will generally lead to cumbersome and non-intuitive expressions.

Instead, we again approach this using eigenoperators. 
Since $H_0 = H_A + H_B$ is a sum of Hamiltonians that live on separate Hilbert spaces, it follows that the eigenoperator decomposition of $V$ can  be written as
\begin{equation}
    V = \sum_{\omega_A,\omega_B} V_{\omega_A+\omega_B},
\end{equation}
where $\omega_A$ runs over all transition frequencies $E_{a'}-E_a$ of $H_A$ and $\omega_B$ runs over all frequencies $E_{b'} - E_b$ of $B$.
To systematically construct $V_{\omega_A+\omega_B}$ one can use the fact that any kind of two-body interaction can always be written as 
\begin{equation}
    V = \sum_\alpha A_\alpha B_\alpha,
\end{equation}
for operators $A_\alpha$ and $B_\alpha$  acting on $A$ and $B$ respectively. If we now decompose each $A_\alpha$ in eigenoperators of $H_A$, and each $B_\alpha$ in eigenoperators of $H_B$, then we can simply add them together to get the eigenoperators of $H_A + H_B$:
\begin{equation}
    V_{\omega_A+\omega_B} = \sum_\alpha A_{\alpha,\omega_A} B_{\alpha,\omega_B}.
\end{equation}

With Fig.~\ref{fig:drawing}(b) in mind, the eigenoperators that should belong to $V_d$ are those for which $\omega_A = 0$, while all eigenoperators with $\omega_A \neq 0$ should belong to $V_o$:
\begin{equation}
    \begin{aligned}
        V_d &= \sum_{\omega_B} V_{\omega_B}, 
        \\[0.2cm]
        V_o &= \sum_{\omega_A \neq 0} \sum_{\omega_B} V_{\omega_A + \omega_B}.
    \end{aligned}
\end{equation}
With these definitions, Eq.~\eqref{Hprime_eigen_final_d} now readily yields 
\begin{equation}
\begin{aligned}
    H' &\simeq H_A + H_B + \sum_{\omega_B} V_{\omega_B} 
    \\[0.2cm]
    &- \sum_{\omega_A, \omega_A'>0}\sum_{\omega_B,\omega_B'} f_{\omega_A-\omega_B, \omega_A'-\omega_B'} [V_{\omega_A-\omega_B}^{}, V_{\omega_A'-\omega_B'}^\dagger].
\end{aligned}
\end{equation}
The sum is over only the positive frequencies of $A$, but is unrestricted over $B$. The reason is because, if $\omega_A>0$ then, by assumption $\omega_A-\omega_B>0$ for any $\omega_B$.
This formula becomes more intuitive once we consider concrete examples.

\subsection{Dispersive system is a  bosonic mode}
\label{sec:dispersive_single_bosonic}

Suppose system $B$ is arbitrary, but system $A$ is composed of a single bosonic mode with $H_A = \omega_r a^\dagger a$ and 
\begin{equation}\label{dispersive_single_b_V}
    V \equiv V_o = a^\dagger B + B^\dagger a,
\end{equation}
where $B$ is an arbitrary operator of system $B$ (which may or may not be Hermitian). 
Decomposing $B = \sum_{\omega_B} B_{\omega_B}$, in eigenoperators of $H_B$, we
then find 
\begin{equation}\label{eqou1098710983712890719871}
    V = \sum_{\omega_B} \big(a^\dagger B_{\omega_B} + B_{\omega_B}^\dagger a\big).
\end{equation}
The set of eigenoperators having positive frequency is therefore $V_{\omega_r-\omega_B} = B_{\omega_B}^\dagger a$ (which is positive because $\omega_r \gg |\omega_B|$ by assumption).
Eq.~\eqref{S_sol} becomes
\begin{equation}
        S = - \sum_{\omega_B} \frac{1}{\omega_r-\omega_B}\Big( B_{\omega_B}^\dagger a - a^\dagger B_{\omega_B}\Big),
\end{equation}
where the sum is over all $\omega_B$, positive, negative or zero.
It is convenient to define a ``renormalized'' $B$ operator as 
\begin{equation}\label{dispersive_renormalized_B}
    B_r = \sum_{\omega_B} \frac{1}{\omega_r-\omega_B}B_{\omega_B},
\end{equation}
so that
\begin{equation}\label{dispersive_S_op_with_Br}
    S = a^\dagger B_r - B_r^\dagger a.
\end{equation}
Next we need to compute 
\begin{equation}
    [S,V_o] = [a^\dagger B_r^{} - B_r^\dagger, a^\dagger B + B^\dagger a].
\end{equation}
To get to the final result~\eqref{Hprime_eigen_final_d} we only need the block off-diagonal part, which means we can discard the commutators with $a^{\dagger 2}$ and $a^2$. 
That is:
\begin{equation}\label{398u41938498137019801981}
\begin{aligned}
    [S,V_o]_d &= [a^\dagger B_r, B^\dagger a]- [B_r^\dagger a, a^\dagger B]
    \\[0.2cm]
    &= -a^\dagger a\left( [B^\dagger,B_r] + [B_r^\dagger,B]\right) - (B^\dagger B_r + B_r^\dagger B)
\end{aligned}    
\end{equation}
Eq.~\eqref{Hprime_final_d} therefore becomes 
\begin{equation}\label{dispersive_single_mode_Hprime}
    \begin{aligned}
        H' =& \left(\omega_r - \tfrac{1}{2}[B^\dagger,B_r] -\tfrac{1}{2}[B_r^\dagger,B]\right) a^\dagger a
        \\[0.2cm]
        &+ H_B - \tfrac{1}{2}(B^\dagger B_r + B_r^\dagger B).
    \end{aligned}
\end{equation}
This is a very useful formula, since we place no assumptions on what system $B$ is. 
The second line describes the effective Hamiltonian of $B$ when mode $a$ is in the vacuum. The first line describes both a renormalization of the frequency of $A$ due to the presence of $B$, as well as the effective Hamiltonian of $B$ whenever $A$ is in a specific Fock state. 
The most usual application of this result is to the dispersive Jaynes-Cummings or Rabi models, which we treat in Sec.~\ref{sec:dispersive_jaynes_cummings}. 

The above calculation also readily extends to the case when we have multiple bosonic modes:
\begin{align}\label{dispersive_multi_boson_V}
    H_0 &= \sum_{i}\omega_{ri} a_i^\dagger a_i^{} + H_B,
    \\[0.2cm]
    V &= V_o = \sum_{i} B_i^\dagger a_i^{} + a_i^\dagger B_i^{}.
    \label{dispersive_multi_boson_V}
\end{align}
The Schrieffer-Wolff Hamiltonian~\eqref{Hprime_final_d} becomes
\begin{equation}\label{dispersive_Hprime_multi_bosonic_modes}
\begin{aligned}
    H' &= \sum_{i} \omega_{ri}a_i^\dagger a_i - \tfrac{1}{2}\sum_{i,j}\left( 
    a_i^\dagger a_j^{} [B_j^\dagger,B_{ri}] + a_j^\dagger a_i^{} [B_{ri}^\dagger, B_j]
    \right)
    \\[0.2cm]
    &\qquad + H_B - \tfrac{1}{2}\sum_{i} \big(B_i^\dagger B_{ri}^{} + B_{ri}^\dagger B_i\big),
\end{aligned}    
\end{equation}
where 
$B_{ri} = \sum_{\omega_{B}} B_{i\omega_{B}}/(\omega_{ri}-\omega_B)$ and $B_{i\omega_B}$ is the eigenoperator decomposition of each $B_i$ with respect to $H_B$. 
The terms with $a_i^\dagger a_j^{}$ and $j\neq i$ will involve energy differences $\omega_{ri}-\omega_{rj}$. They should therefore only be kept if the bosonic frequencies are close to one another. Otherwise, one may keep only the $i=j$ contributions.
In Sec.~\ref{sec:quadratic_fermions_linear_bosons} we apply this formula to cubic fermion-boson interactions, such as the electron-phonon coupling. 

\subsubsection{Dispersive Jaynes-Cummings and Rabi models}\label{sec:dispersive_jaynes_cummings}

Suppose system $B$ is a qubit with 
\begin{equation}
    H_B = \frac{\omega_q}{2}\sigma_z,
    \qquad 
    B = g \sigma_-,
\end{equation}
which is the Jaynes-Cummings model. 
This operator $B$ is already an eigenoperator of $H_B$, so that the renormalized operator $B_r$ in Eq.~\eqref{dispersive_renormalized_B} will be simply 
\begin{equation}
    B_r = \tfrac{g}{\omega_r-\omega_q}\sigma_-.
\end{equation}
Eq.~\eqref{dispersive_single_mode_Hprime} then becomes the familiar dispersive Jaynes-Cummings formula
\begin{equation}
    H' = \Big(\omega_r - \chi \sigma_z\Big) a^\dagger a + \frac{(\omega_q-\chi)}{2} \sigma_z  -\frac{\chi}{2}.
\end{equation}
where $\chi = \tfrac{g^2}{\omega_r - \omega_0}$ is the dispersive/Lamb shift.

Instead, suppose we had a Rabi model 
\begin{equation}
    H_B = \frac{\omega_q}{2}\sigma_z,
    \qquad 
    B = g \sigma_x.
\end{equation}
In this case 
\begin{equation}
    B_r = \tfrac{g}{\omega_r-\omega_q} \sigma_- + \tfrac{g}{\omega_r+\omega_q}\sigma_+.
\end{equation}
But because $\sigma_-^2 = 0$, Eq.~\eqref{dispersive_single_mode_Hprime} ends up yielding a similar expression 
\begin{equation}
    H' = \Big( \omega_r - \tilde{\chi}\Big)a^\dagger a + \frac{(\omega_q - \tilde{\chi})}{2} \sigma_z - \frac{g^2\omega_r}{\omega_r^2-\omega_q^2},
\end{equation}
where $\tilde{\chi} = \tfrac{g^2}{\omega_r - \omega_q} - \tfrac{g^2}{\omega_r + \omega_q}$.
So for this particular model, both interactions lead to the same dispersive model, except for a  change in the dispersive shift.

Let us also consider the generalization of the Jaynes-Cummings model to the case where the system is a qudit, with $d$ levels. 
We take $H_B = \sum_j E_j \sigma_j$ and assume the energies $E_j$, and the transition frequencies $E_i - E_j$, are  all non-degenerate. 
We also assume, for illustration purposes, that Eq.~\eqref{dispersive_single_b_V} has 
$B = \sum_j b_j \sigma_{j,j+1}$, for coefficients $b_j$; i.e., the operator $B$ only acts between neighboring energy levels of $H_B$. 
In this case Eq.~\eqref{dispersive_renormalized_B} becomes
\begin{equation}
    B_r = \sum_j \frac{b_j}{\omega_r - \omega_{j+1,j}} \sigma_{j,j+1},
\end{equation}
where $\omega_{j+1,j} = E_{j+1}-E_j$. 
We then get 
\begin{equation}
\begin{aligned}
    B^\dagger B_r &= B_r^\dagger B = \sum_j \chi_j \sigma_{j+1}, 
    \\[0.2cm]
    B_r B^\dagger &= B B_r^\dagger = \sum_j \chi_j \sigma_j,
\end{aligned}    
\end{equation}
where $\chi_j = \tfrac{|b_j|^2}{\omega_r - \omega_{j+1,j}}$. 
Thus Eq.~\eqref{dispersive_single_mode_Hprime} becomes
\begin{equation}
    H' = \left( \omega_r - \sum_j \chi_j (\sigma_{j+1}-\sigma_j)\right)a^\dagger a + \sum_j (E_j-\chi_j) \sigma_{j+1}.
\end{equation}
The last term describes how the frequency of each level $\sigma_{j+1} = |j+1\rangle\langle j+1|$ is renormalized by a factor $\chi_j$ due to the interaction with the bosonic mode.
This formula is widely used in the superconducting circuits literature, c.f.~Ref.~\cite{blais2021}.

\subsection{Dispersive system is a  qubit}\label{sec:dispersive_single_qubit}

Consider again a general dispersive scenario, where system $B$ can be anything. 
However, suppose that instead of $A$ being a  bosonic mode, as in Sec.~\ref{sec:dispersive_single_bosonic}, it is a  qubit with
\begin{equation}
    H_A = \tfrac{\omega_r}{2}\sigma_z, 
    \qquad 
    V = B^\dagger \sigma_- + \sigma_+ B.
\end{equation}
The same logic applies in this case, leading to
\begin{equation}
    S = \sigma_+ B_r - B_r^\dagger \sigma_+,
\end{equation}
with $B_r$ still given in Eq.~\eqref{dispersive_renormalized_B}.
The block diagonal part of $[S,V_o]$ is then 
\begin{equation}
    \begin{aligned}
        [S,V_o]_d &= [\sigma_+ B_r, B^\dagger \sigma_-] - [B_r^\dagger \sigma_-, \sigma_+ B]
        \\[0.2cm]
        &= \left(B_r^\dagger B+ B B_r^\dagger \right)\sigma_+\sigma_-
        - \left(B^\dagger B_r + B_r^\dagger B \right)\sigma_-\sigma_+.
    \end{aligned}
\end{equation}
Hence
\begin{equation}\label{dispersive_qubit_Hprime}
    \begin{aligned}
        H' =& H_B + \tfrac{1}{2}\left(\omega_r+ 
        B_r B^\dagger + B B_r^\dagger \right)\sigma_+ \sigma_- 
        \\[0.2cm]
        &- \tfrac{1}{2}\left(\omega_r +
        B^\dagger B_r + B_r^\dagger B
        \right)\sigma_-\sigma_+.
    \end{aligned}
\end{equation}
We therefore see that there will be 
two effective Hamiltonians for system $B$, depending on whether the spin is up or down: 
\begin{equation}
    H_B' = \begin{cases}
        H_B + \tfrac{1}{2}(B_r B^\dagger + B B_r^\dagger) & \text{spin up}
        \\[0.2cm]
        H_B +\tfrac{1}{2}(B^\dagger B_r + B_r^\dagger B) & \text{spin down},
    \end{cases}
\end{equation}
where we neglected a constant $\pm \omega_r/2$. 

\subsubsection{Tight-binding chain dispersively coupled to a single qubit}\label{sec:dispersive_tight_binding_qubit}

As an illustration of Eq.~\eqref{dispersive_qubit_Hprime},  consider a tight-binding chain (of either fermions or bosons) dispersively coupled to a single qubit:
\begin{equation}
\begin{aligned}
    H_0 &= \tfrac{\omega_r}{2}\sigma_z + \sum_k \epsilon_k^{} b_k^\dagger b_k^{},
    \\[0.2cm]
    V&= B^\dagger \sigma_- + \sigma_+ B,\qquad 
    B = \sum_k g_k b_k.
\end{aligned}    
\end{equation}
This kind of coupling allows for the qubit to couple to multiple modes $b_k$. 
This is widely used e.g., in the context of giant atoms~\cite{roccati2024,roccati2024a}, which couple to multiple lattice sites. 
Eq.~\eqref{dispersive_renormalized_B} gives $B_r = \sum_k \tfrac{g_k}{\omega_r-\epsilon_k}b_k$.
Therefore 
\begin{equation}
\begin{aligned}
    B_r^\dagger B + B^\dagger B_r &= \sum_{k,q} g_k^* g_q^{} \left(\frac{1}{\omega_r - \epsilon_k}+\frac{1}{\omega_r - \epsilon_q} \right)b_k^\dagger b_q^{},
    \\[0.2cm]
    B_r B^\dagger + B B_r^\dagger &= \sum_k \frac{2|g_k|^2}{\omega_r-\epsilon_k} \mp (B_r^\dagger B + B^\dagger B_r),
\end{aligned}    
\end{equation}
with the minus sign for fermions and the plus sign for bosons.

This can now be plugged in Eq.~\eqref{dispersive_qubit_Hprime}.
We treat fermions and bosons separately. 
If the chain is made up of fermions, the effective Hamiltonian will decouple completely the qubit and chain components: 
\begin{equation}
    \begin{aligned}
        H' =&  \frac{\omega_r}{2}\sigma_z + \left( \sum_k \frac{|g_k|^2}{\omega_r-\epsilon_k}\right)\sigma_+ \sigma_- 
        \\[0.2cm]
        +& \sum_k \epsilon_k^{} b_k^\dagger b_k^{} - \tfrac{1}{2}\sum_{k,q} g_k^* g_q^{} \left(\frac{1}{\omega_r - \epsilon_k}+\frac{1}{\omega_r - \epsilon_q} \right)b_k^\dagger b_q^{}.
    \end{aligned}
\end{equation}
The presence of the qubit modifies the fermionic Hamiltonian, but the modification is independent of the spin's state. 

Conversely, if the chain is bosonic we get 
\begin{equation}
    \begin{aligned}
        H' =&  \frac{\omega_r}{2}\sigma_z + \left( \sum_k \frac{|g_k|^2}{\omega_r-\epsilon_k}\right)\sigma_+ \sigma_- 
        \\[0.2cm]
        +& \sum_k \epsilon_k^{} b_k^\dagger b_k^{} + \tfrac{1}{2}\sigma_z\sum_{k,q} g_k^* g_q^{} \left(\frac{1}{\omega_r - \epsilon_k}+\frac{1}{\omega_r - \epsilon_q} \right)b_k^\dagger b_q^{}.
    \end{aligned}
\end{equation}
To understand the significance of the last term, notice that the leading order contributions  correspond to the terms with $q=k$.
Retaining only these terms yields for the bosonic chain an effective Hamiltonian 
\begin{equation}
    H_B' \simeq \sum_k \left( 
    \epsilon_k \pm \frac{|g_k|^2}{\omega_r - \epsilon_k} 
    \right) b_k^\dagger b_k^{},
\end{equation}
with the plus or minus sign depending on whether the spin is up or down. 
By controlling the spin value, one may therefore affect the chain's dispersion relation, and hence how excitations propagate through the chain.

\section{Additional examples} 

\subsection{Two dispersively-coupled tight-binding chains}\label{sec:tight_binding}

We consider here a dispersive problem (Fig.~\ref{fig:drawing}(b)) where $H_A$, $H_B$ and $V$ are all quadratic:
\begin{equation}\label{tight_binding_HA_HB_V}
\begin{aligned}
    H_A &= \sum_i E_i a_i^\dagger a_i, 
    \\[0.2cm]
    H_B &= \sum_k \epsilon_k b_k^\dagger b_k,
    \\[0.2cm]
    V &= \sum_{i,k}\Big(v_{ik}^{} b_k^\dagger a_i
    +
    v_{ik}^* a_i^\dagger b_k^{}\Big),
\end{aligned}    
\end{equation}
for generic coefficients $v_{ik}$.
We assume $a_i$ and $b_k$ can be either fermionic or bosonic, and consider all 4 possible combinations. 

Each term $v_{ik}^{} b_k^\dagger a_i^{}$ is an eigenoperator of $H_A+H_B$ with frequency $E_i-\epsilon_k$.
Which of these belong to $V_d$ or $V_o$ will depend on the problem at hand. 
Below we assume that all terms belong to $V_o$. If one wishes to include some of them in $V_d$, it suffices to exclude these terms from the sums. 
The transformation $S$, from Eq.~\eqref{S_sol}, reads 
\begin{equation}
    S = - \sum_{i,k} \frac{v_{ik}}{E_i-\epsilon_k}\big(b_k^\dagger a_i^{} - a_i^\dagger b_k^{}\big).
\end{equation}
The SW Hamiltonian is, from Eq.~\eqref{Hprime_eigen_final_d},
\begin{equation}
\begin{aligned}
    H' &= H_A + H_B - \sum_{i,j,k,q} f_{ij}^{kq} [b_k^\dagger a_i^{}, a_j^\dagger b_q^{}]
    \\[0.2cm]
    &=H_A + H_B - \sum_{i,j,k,q} f_{ij}^{kq}\Big(a_i^{} a_j^\dagger b_k^\dagger b_q^{} - a_j^\dagger a_i^{} b_q^{} b_k^\dagger\Big),
\end{aligned}    
\end{equation}
where 
\begin{equation}
    f_{ij}^{kq} = \frac{v_{ik}^{} v_{jq}^*}{2}\bigg( \frac{1}{E_i-\epsilon_k} + \frac{1}{E_j-\epsilon_q}\bigg).
\end{equation}
The physics that ensues from this will depend on the algebra of the $a$ and $b$ operators:
\begin{itemize}\itemsep0cm
    \item If $a_i$ and $b_k$ are of the same species  (both fermionic or both bosonic) we get     \begin{equation}\label{tight_binding_same_species}
        H' = \sum_{i,j} \Big( E_i\delta_{i,j} + \sum_k f_{ij}^{kk}\Big) a_j^\dagger a_i^{}
        + \sum_{k,q} \Big( \epsilon_k \delta_{k,q} - \sum_i f_{ii}^{kq} \Big) b_k^\dagger b_q^{},
\end{equation}
This has the form $H_A' + H_B'$, so that the transformation fully decouples $A$ and $B$. 
\item If $a_i$ are fermions but $b_k$ are bosons we get an extra term 
\begin{equation}  
\begin{aligned}
        H' =& \sum_{i,j} \Big( E_i\delta_{i,j} + \sum_k f_{ij}^{kk}\Big) a_j^\dagger a_i^{}
        + \sum_{k,q} \Big( \epsilon_k \delta_{k,q} - \sum_i f_{ii}^{kq} \Big) b_k^\dagger b_q^{}
        \\[0.2cm]
        &+ 2 \sum_{i,j,k,q} f_{ij}^{kq} a_j^\dagger a_i^{} b_k^\dagger b_q^{}.
\end{aligned}        
\end{equation}
If $a_i$ are bosons but $b_k$ are fermions, we get the same extra term, but with an opposite sign. 
\end{itemize}


As an example of Eq.~\eqref{tight_binding_same_species} suppose $a_i$ and $b_k$ are fermionic operators representing two energy bands in a crystal. Suppose the coupling is resonant, $v_{ik} = v \delta_{ik}$ and that the spacing between the two bands is also constant, $E_i = \epsilon_i + \Delta$. Then 
\begin{equation}
    f_{ij}^{kq} = \delta_{ik}\delta_{jq}\frac{|v|^2}{\Delta},
\end{equation}
and so 
\begin{equation}
    H' = \sum_i (E_i + |v|^2/\Delta)a_i^\dagger a_i^{} + \sum_i (\epsilon_i - |v|^2/\Delta)b_i^\dagger b_i^{}.
\end{equation}
The interaction is therefore translated into an effective repulsion between the two bands.

\subsection{Cubic fermion-boson coupling}\label{sec:quadratic_fermions_linear_bosons}

Next we consider a fermionic system coupled dispersively to a set of bosonic modes, such as an optical cavity, or phonons in a crystal. 
The Hamiltonian is taken as 
\begin{equation}\label{ceren_H0V_b_ops}
    \begin{aligned}
        H_0 &= \sum_{i} \omega_{ri}a_i^\dagger a_i + \sum_{n,m} h_{nm} b_n^\dagger b_m^{}
        \\[0.2cm]
        V &= \sum_{i,n,m} b_n^\dagger b_m^{} \Big(M_{inm}^{} a_i^\dagger + M_{imn}^* a_i^{}\Big),
    \end{aligned}
\end{equation}
where $\omega_{ri}$ are the energies of the bosonic modes and $h=h^\dagger$ is a matrix representing the fermionic system. 
Finally, $M_{inm}$ are the interaction strengths. 
This kind of cubic interaction is typical, as it is the simplest form which preserves the fermionic particle number.

To proceed we first diagonalize the fermionic Hamiltonian by diagonalizing the matrix $h$ as $h = R \epsilon R^\dagger$, where $\epsilon$ is a diagonal matrix containing the eigenenergies, and $R$ is unitary. 
Let $\bm{b}$ denote a vector containing the fermionic operators $b_n$. 
Introducing a new set of fermionic operators as $\bm{c} = R^\dagger \bm{b}$ (or $\bm{b} = R \bm{c}$), we then get 
\begin{equation}
\begin{aligned}\label{ceren_V_cca}
    H_B &= \bm{b}^\dagger h \bm{b} = \bm{c}^\dagger \epsilon \bm{c} = \sum_k \epsilon_k c_k^\dagger c_k^{},
    \\[0.2cm]
    V &= \sum_{i,k,q}  c_k^\dagger c_q^{} \Big(\mathcal{M}_{ikq}^{} a_i^\dagger + \mathcal{M}_{iqk}^* a_i^{}\Big),
\end{aligned}
\end{equation}
where 
\begin{equation}\label{49837198316781687}
    \mathcal{M}_i = R^\dagger M_i R. 
\end{equation}
We introduce both the non-diagonal model~\eqref{ceren_H0V_b_ops} and the diagonal model~\eqref{ceren_V_cca} since each may be more convenient depending on the application. 

In the language of Eq.~\eqref{dispersive_multi_boson_V}, the  fermionic operators $B_i$, coupling to each bosonic mode, are 
\begin{equation}
    B_i = \sum_{n,m} M_{inm} b_n^\dagger b_m^{} = \bm{b}^\dagger M_i \bm{b} = \bm{c}^\dagger \mathcal{M}_i \bm{c}, 
\end{equation}
with one operator $B_i$ for each bosonic mode present. 
Because this is now written in terms of the modes $c_k$ that diagonalize $H_B$, its eigenoperator decomposition can be readily seen to be
\begin{equation}
    B_i = \sum_{k,q} B_{i,\epsilon_q-\epsilon_k},
    \qquad 
    B_{i,\epsilon_q-\epsilon_k} = \mathcal{M}_{ikq} c_k^\dagger c_q.
\end{equation}
The renormalized operators $B_{ri}$ in Eq.~\eqref{dispersive_renormalized_B} then read 
\begin{equation}
    B_{ri} = \sum_{k,q} \frac{\mathcal{M}_{ikq}}{\omega_{ri} - \epsilon_q + \epsilon_k} c_k^\dagger c_q.
\end{equation}
The effective Hamiltonian of the fermions, assuming that the bosonic modes are in the vacuum,  immediately follow  from the second line of Eq.~\eqref{dispersive_Hprime_multi_bosonic_modes}:
\begin{equation}\label{ceren_bosons_HBp}
    H_B' = \sum_k\epsilon_k c_k^\dagger c_k^{} + \sum_{k,k',p,p'} \mathcal{A}_{kk'}^{pp'} c_{k'}^\dagger c_k^{} c_p^\dagger c_{p'}^{},
\end{equation}
with coefficients 
\begin{equation}\label{ceren_A_tensor}
    \mathcal{A}_{kk'}^{pp'} = -\tfrac{1}{2}\sum_i \mathcal{M}_{ikk'}^* \mathcal{M}_{ipp'} \left(
    \frac{1}{\omega_{ri} +\epsilon_k-\epsilon_{k'}}
    +\frac{1}{\omega_{ri} +\epsilon_p-\epsilon_{p'}}
    \right).
\end{equation}

To gain further insights, we next  discuss two alternative ways of writing Eq.~\eqref{ceren_bosons_HBp}, that may be more convenient depending on the application.
First, using the fermionic commutation relations we can write 
\begin{equation}
\begin{aligned}\label{ceren_bosons_HBp2}
    H_B' =& \sum_k \epsilon_k c_k^\dagger c_k^{} + \sum_{k,p} \mathcal{G}_{kp} c_k^\dagger c_p^{} 
    +\sum_{k,k',p,p'} \mathcal{F}_{kk'}^{pp'} c_{k'}^\dagger c_p^\dagger c_k^{} c_{p'}^{},
\end{aligned}    
\end{equation}
where 
\begin{equation}\label{ceren_bosons_HBp2_coefficients}
    \begin{aligned}
        \mathcal{G}_{kp} &= -\tfrac{1}{2}\sum_i \sum_q \mathcal{M}_{iqk}^*\mathcal{M}_{iqp}\left( 
        \frac{1}{\omega_{ri}+\epsilon_q - \epsilon_k}
        +
        \frac{1}{\omega_{ri}+\epsilon_q - \epsilon_p}
        \right),
        \\[0.2cm]
        \mathcal{F}_{kk'}^{pp'} &= \tfrac{1}{2}\sum_i \bigg\{ \frac{\mathcal{M}_{ikk'}^* \mathcal{M}_{ipp'}}{\omega_{ri} + (\epsilon_k - \epsilon_{k'})}
        +
        \frac{\mathcal{M}_{ik'k} \mathcal{M}_{ip'p}^*}{\omega_{ri} -( \epsilon_k - \epsilon_{k'})}      
        \bigg\}.
    \end{aligned}
\end{equation}
If the perturbation $V$ in Eq.~\eqref{ceren_V_cca} is of the form 
$V = \sum_{i,k,q}  \mathcal{M}_{ikq}^{}c_k^\dagger c_q^{} \Big( a_i^\dagger + a_i^{}\Big)$ 
then the $\mathcal{M}_{ikq}$
must be Hermitian in the fermionic indices; i.e., $\mathcal{M}_{ikq}^* = \mathcal{M}_{iqk}^*$.
In this case the coefficients $\mathcal{F}_{kk'}^{pp'}$ 
in Eq.~\eqref{ceren_bosons_HBp2_coefficients}
simplify further to 
\begin{equation}
    \mathcal{F}_{kk'}^{pp'} = \sum_i \mathcal{M}_{ikk'}^* \mathcal{M}_{ipp'}^{}
    \frac{\omega_{ri}}{\omega_{ri}^2-(\epsilon_k-\epsilon_{k'})^2}
\end{equation}

Second, under the assumption that the interaction is dispersive, $\omega_{ri} \gg |\epsilon_k-\epsilon_{k'}|$, one may wish to further expand 
\begin{equation}
    \mathcal{A}_{kk'}^{pp'} \simeq -\sum_i\frac{\mathcal{M}_{ikk'}^* \mathcal{M}_{ipp'}}{\omega_{ri}} + \tfrac{1}{2} \sum_i 
    \frac{\mathcal{M}_{ikk'}^* \mathcal{M}_{ipp'}}{\omega_{ri}^2}(\epsilon_k-\epsilon_{k'} + \epsilon_p -\epsilon_{p'}).
\end{equation}
This will, of course, lead to a worse approximation than Eq.~\eqref{ceren_bosons_HBp}. 
But it also makes some of the underlying structure clearer, since Eq.~\eqref{ceren_bosons_HBp} now takes the more compact form
\begin{equation}
    \begin{aligned}
        H_B' =& \bm{c}^\dagger \epsilon \bm{c} - \sum_i 
        \frac{1}{\omega_{ri}} (\bm{c}^\dagger \mathcal{M}_i^\dagger \bm{c})(\bm{c}^\dagger \mathcal{M}_i \bm{c})
        \\[0.2cm]+\sum_i  &
        \frac{1}{2\omega_{ri}^2}\bigg\{ 
        (\bm{c}^\dagger \mathcal{M}_i^\dagger \bm{c}) (\bm{c}^\dagger [\epsilon, \mathcal{M}_i] \bm{c}) 
        -(\bm{c}^\dagger [\epsilon, \mathcal{M}_i]^\dagger \bm{c}) (\bm{c}^\dagger \mathcal{M}_i \bm{c}) 
       \bigg\}
    \end{aligned}
\end{equation}
The dominant contribution is in the first line.
The convenient aspect of this formula is that it is easier to  rewrite it in terms of the original (pre-diagonalization) operators $b_n$ [Eq.~\eqref{ceren_H0V_b_ops}], which might have a more direct physical interpretation. 
Using $\bm{c} = R^\dagger \bm{b}$ we get 
\begin{equation}\label{ceren_bosons_HBp3}
    \begin{aligned}
        H_B' =& \bm{b}^\dagger h \bm{b} - \sum_i 
        \frac{1}{\omega_{ri}} (\bm{b}^\dagger M_i^\dagger \bm{b})(\bm{b}^\dagger M_i \bm{b})
        \\[0.2cm]+\sum_i  &
        \frac{1}{2\omega_{ri}^2}\bigg\{ 
        (\bm{b}^\dagger M_i^\dagger \bm{b}) (\bm{b}^\dagger [h, M_i] \bm{b}) 
        -(\bm{b}^\dagger [h,M_i^\dagger] \bm{b}) (\bm{b}^\dagger M_i \bm{b}) 
       \bigg\},
    \end{aligned}
\end{equation}
where we also used Eq.~\eqref{49837198316781687} and $h = R \epsilon R^\dagger$. 
Again, the leading order contribution here is really the first line
\begin{equation}\label{ceren_bosons_HBp3_simple}
    H_B' = \bm{b}^\dagger h \bm{b} - \sum_i 
        \frac{1}{\omega_{ri}} (\bm{b}^\dagger M_i^\dagger \bm{b})(\bm{b}^\dagger M_i \bm{b}),
\end{equation}
which provides a simple and easy to apply expression.

\subsubsection{Electron-phonon interaction}\label{sec:electron_phonon}

An example application of Eq.~\eqref{ceren_bosons_HBp2} is to the Fr\"olich Hamiltonian describing the interaction of electrons with phonons~\cite{frohlich1952}. 
In this case: 
\begin{equation}    
        c_k \to c_{\bm{k}\sigma}, 
        \qquad 
        a_i \to a_{\bm{p}},    
\end{equation}
where $\bm{k},\bm{p}$ are momenta and $\sigma=\pm1$ is the spin.
The interaction coefficients $\mathcal{M}_{ikq}$ in Eq.~\eqref{ceren_V_cca} are replaced with the spin- and momentum-conserving form
\begin{equation}
    \mathcal{M}_{\bm{p},\bm{k}\sigma, \bm{k}'\sigma'} = \delta_{\sigma,\sigma'} \delta_{\bm{p},\bm{k}-\bm{k}'} g_{\bm{p}},
\end{equation}
with coefficients $g_{\bm{p}}$. 
The quartic term in Eq.~\eqref{ceren_bosons_HBp2} then becomes the BCS Hamiltonian
\begin{equation}
    \sum_{\bm{k},\bm{q},\bm{p},\sigma,\sigma'} \frac{|g_{\bm{p}}|^2 \omega_{\bm{p}}}{\omega_{\bm{p}}^2 - (\epsilon_{\bm{k},\sigma}-\epsilon_{\bm{k}-\bm{p},\sigma})^2} c_{\bm{k}-\bm{p},\sigma}^\dagger c_{\bm{q},\sigma'}^\dagger c_{\bm{k},\sigma} c_{\bm{q}-\bm{p}, \sigma}.
\end{equation}
We derived this result under the assumption that the phonons are the dispersive element; i.e, $\omega_{\bm{p}} \gg |\epsilon_{\bm{k}, \sigma} - \epsilon_{\bm{k'},\sigma'}|$. 
In this case the potential is always positive and therefore the phonons mediate a repulsive interaction between the electrons.
However, one would also arrive at the same result if the dispersive elements were the fermions instead; and in this case the interaction would be negative, which is the ingredient required for superconductivity. 

\subsubsection{Graphene in a cavity}\label{sec:graphene}

As a second example, now focusing on the form~\eqref{ceren_bosons_HBp3}, we study 
a problem recently investigated in~\cite{dag2023}, which consists of a graphene sheet placed inside an optical cavity.
In this case: 
\begin{equation}
    c_k \to c_{A\bm{k}}, ~c_{B\bm{k}},
    \qquad 
    a_i \to a_L,~a_R,
\end{equation}
where $A/B$ labels the two graphene sublattices and $a_L$ and $a_R$ are the left and right circularly polarized modes of the cavity.
The unperturbed Hamiltonian near the $\bm{K}$ valley is 
\begin{equation}
        H_0 =  v_F\sum_{\bm{k}}\Big\{(k_x+ik_y) c_{A\bm{k}}^\dagger c_{B\bm{k}}^{} + (k_x-ik_y) c_{B\bm{k}}^\dagger c_{A\bm{k}}^{} \Big\} + \sum_{i=L,R} \omega_{ri} a_i^\dagger a_i^{},
\end{equation}
where $v_F$ is the Fermi velocity.
In the language of Eq.~\eqref{ceren_H0V_b_ops} (using operators $c_{A/B,\bm{k}}$ instead of $b_n$) this Hamiltonian corresponds to a matrix 
\begin{equation}
    h_{\alpha \bm{k}, \beta\bm{q}} = \delta_{\bm{k},\bm{q}} \tilde{h}_{\alpha,\beta}^{\bm{k}},
    \qquad 
    \tilde{h}^{\bm{k}} = v_F\begin{pmatrix}
        0 & k_x+ik_y \\ k_x - i k_y & 0
    \end{pmatrix},
\end{equation}
where $\alpha,\beta = A,B$ labels the sublattices.
Notice that we can also write $\tilde{h}^{\bm{k}} = k_x \sigma_x -  k_y \sigma_y$, 
where $\sigma_\mu$ are Pauli matrices.

The cavity-graphene interaction, as shown by the authors of Ref.~\cite{dag2023}, can be written as 
\begin{equation}
\begin{aligned}
    V &= -v_F \sum_{\bm{k}} \Big\{ g_L \Big(c_{A\bm{k}}^\dagger c_{B\bm{k}}^{}a_L + 
    c_{B\bm{k}}^\dagger c_{A\bm{k}}^{}a_L^\dagger\Big) 
    \\[0.2cm]
    &\qquad + 
    g_R \Big(c_{A\bm{k}}^\dagger c_{B\bm{k}}^{}a_R^\dagger + 
    c_{B\bm{k}}^\dagger c_{A\bm{k}}^{}a_R\Big)\Big\}.
\end{aligned}    
\end{equation}
Thus, in the language of Eq.~\eqref{ceren_H0V_b_ops} the interaction coefficients are 
\begin{equation}
\begin{aligned}
    M_{L, \alpha\bm{k}, \beta\bm{q}} &= - v_Fg_L \delta_{\bm{k},\bm{q}} 
    (\sigma_+)_{\alpha\beta} 
    \\[0.2cm]
    M_{R, \alpha\bm{k}, \beta\bm{q}} &= - v_Fg_R \delta_{\bm{k},\bm{q}} (\sigma_-)_{\alpha\beta}
\end{aligned}    
\end{equation}

With these definitions we can now immediately apply Eq.~\eqref{ceren_bosons_HBp3}. 
We have: 
\begin{equation*}
\begin{aligned}
    \bm{b}^\dagger M_L \bm{b} &\to -v_F \sum_{\bm{k}} g_L c_{A\bm{k}}^\dagger c_{B\bm{k}}^{}
    \\[0.2cm]
    \bm{b}^\dagger M_R \bm{b} &\to -v_F \sum_{\bm{k}} g_R c_{B\bm{k}}^\dagger c_{A\bm{k}}^{}
    \\[0.2cm]
    \bm{b}^\dagger[h,M_L]\bm{b} &\to -v_F\sum_{\bm{k}} g_L\Big\{(k_x-i k_y) (n_{B\bm{k}} -
    n_{A\bm{k}})\Big\},
    \\[0.2cm]
    \bm{b}^\dagger[h,M_R]\bm{b} &\to -v_F\sum_{\bm{k}} g_R\Big\{(k_x+i k_y) (n_{A\bm{k}} -
    n_{B\bm{k}})\Big\},
\end{aligned}
\end{equation*}
where $n_{\alpha\bm{k}} = c_{\alpha \bm{k}}^\dagger c_{\alpha \bm{k}}^{}$.
Thus the effective fermionic Hamiltonian, when the cavity is in the vacuum is 
\begin{equation}
    \begin{aligned}
        H_B' =& v_F\sum_{\bm{k}}\Big\{(k_x+ik_y) c_{A\bm{k}}^\dagger c_{B\bm{k}}^{} + (k_x-ik_y) c_{B\bm{k}}^\dagger c_{A\bm{k}}^{} \Big\} 
        \\[0.2cm]
        &-  \sum_{\bm{k},\bm{k}'} 
        \Big\{ 
        \frac{v_F^2g_L^2}{\omega_{rL}}c_{B\bm{k}}^\dagger c_{A\bm{k}}^{} c_{A\bm{k}'}^\dagger c_{B\bm{k}'}^{}
        - \frac{v_F^2g_R^2}{\omega_{rR}} c_{A\bm{k}}^\dagger c_{B\bm{k}}^{} c_{B\bm{k}'}^\dagger c_{A\bm{k}'}^{}
        \Big\}
        \\[0.2cm]
        &+ \frac{v_F^2g_L^2}{2\omega_{rL}^2} \sum_{\bm{k},\bm{k}'}c_{B\bm{k}}^\dagger c_{A\bm{k}}^{} (k_x'-ik_y')(n_{B\bm{k}'}-n_{A\bm{k}'}) + \text{h.c.}
        \\[0.2cm]
        &+ \frac{v_F^2g_R^2}{2\omega_{rR}^2} \sum_{\bm{k},\bm{k}'}c_{A\bm{k}}^\dagger c_{B\bm{k}}^{} (k_x'+ik_y')(n_{A\bm{k}'}-n_{B\bm{k}'}) + \text{h.c.}.
    \end{aligned}
\end{equation}
The terms proportional to $1/\omega_{ri}$ describe momentum changes $\bm{k}'-\bm{k}$ in sublattice $A$, and $\bm{k}-\bm{k}'$ in $B$, and vice-versa. 
Conversely, the terms proportional to $1/\omega_{ri}^2$ describe occupation-mediated transfers from sublattice $A$ to $B$, and vice-versa.

\subsection{Non-relativistic limit of the Dirac equation}\label{sec:Dirac}

Around 15 years before the original Schrieffer-Wolff paper, the same technique had already been used by Foldy and Wouthuysen to derive the non-relativistic limit of the Dirac equation~\cite{foldy1950}. 
Consider the Dirac Hamiltonian for a particle in the presence of a (time-independent for simplicity) external electromagnetic field 
\begin{equation}\label{Dirac_H}
    H = \beta m - e \phi + \bm{\alpha} \cdot (\bm{p}- e \bm{A}). 
\end{equation}
Here $\phi(\bm{r})$ and $\bm{A}(\bm{r})$ are the scalar and vector potentials of the field, $m$ is the mass of the particle, $\bm{p}$ its momentum and $\beta, \bm{\alpha}$ a set of $4\times4$ Dirac matrices. For concreteness, one may take e.g. 
\begin{equation}
    \beta = \begin{pmatrix}
        1 & 0 \\ 0 & - 1
    \end{pmatrix},
    \qquad 
    \alpha_i = \begin{pmatrix}
        0 & \sigma_i \\ \sigma_i & 0 
    \end{pmatrix},
\end{equation}
with $\sigma_i$ being the usual Pauli matrices. 
Notice that we can write $\beta = \sigma_z\otimes 1$ and 
$\alpha_i = (\sigma_+ +\sigma_-)\otimes \sigma_i$ (we will use these forms below).

The non-relativistic limit of the Dirac Hamiltonian corresponds to the case where $p \ll m$. Therefore, we may split Eq.~\eqref{Dirac_H} as 
\begin{equation}
    \begin{aligned}
        H_0 &= \beta m, 
        \\[0.2cm]
        V &= - e \phi + \bm{\alpha} \cdot (\bm{p}- e \bm{A}).
    \end{aligned}
\end{equation}
It readily follows that $[H_0, -e\phi] = 0$, so that $-e\phi$ is an eigenoperator of $H_0$ with zero frequency. 
This will therefore belong to $V_d$.
To decompose the other term in eigenoperators we introduce 
\begin{equation}
    V_\pm = \sigma_\pm \otimes \big[\bm{\sigma}\cdot (\bm{p}-e\bm{A})\big].
\end{equation}
The splitting is such that $V_-+V_+ = \bm{\alpha} \cdot (\bm{p}- e \bm{A})$. 
And using the properties of $\beta$ and $\alpha_i$ one may verify that 
\begin{equation}
    [H_0, V_-] = - 2m V_-,
\end{equation}
and $V_+ = V_-^\dagger$. Thus, $V_-$ is the eigenoperator decomposition of the perturbation, with transition frequency $2m$. 
The block off-diagonal part of the perturbation is thus $V_o = V_-+V_+$.

The unitary transformation matrix now follows directly from Eq.~\eqref{S_sol}:
\begin{equation}
    S = - \frac{1}{2m}(V_--V_+) = \frac{\beta}{2m}\bm{\alpha} \cdot (\bm{p}- e \bm{A}).
\end{equation}
Moreover, the SW Hamiltonian follows from Eq.~\eqref{Hprime_eigen_final_d}:
\begin{equation}
\begin{aligned}
    H' &= \beta m - e \phi - \frac{1}{2m}[V_-,V_+].
\end{aligned}    
\end{equation}
Opening up the commutator and noticing that $(\sigma_+\sigma_--\sigma_-\sigma_+)\otimes \bullet = \beta \bullet$ we get
\begin{equation}
    H' = \beta m - e \phi + \frac{\beta}{2m}
    \big[\bm{\sigma}\cdot (\bm{p}-e\bm{A})\big]^2.
\end{equation}
Finally, expanding out the square term, using various  Pauli matrix identities as well as the relation $[p_i, f(\bm{r})] = -i \hbar \partial_i f$, we get 
\begin{equation}
    H' = \beta m - e \phi +\frac{\beta}{2m} (\bm{p}-e\bm{A})^2 - \frac{e\hbar\beta}{2m} (\bm{\sigma}\cdot\bm{B}),
\end{equation}
where $\bm{B} = \nabla \times \bm{A}$ is the magnetic field. 
The entire Hamiltonian is now proportional to $\beta$, and therefore essentially splits symmetrically into two subspaces, each of dimension $2$. 
In the upper subspace, corresponding to the block of $\beta$ having eigenvalue $1$, the effective Hamiltonian reads
\begin{equation}
    H_+' = \frac{1}{2m}(\bm{p}-e\bm{A})^2 - e \phi - \frac{e\hbar}{2m}\bm{\sigma}\cdot\bm{B},
\end{equation}
which is the Schr\"odinger-Pauli Hamiltonian  describing non-relativistic spin 1/2 particles (with gyromagnetic ratio $g \equiv 2$). 
Interestingly, therefore, we see that the non-relativistic Schr\"odinger equation is obtained as a block diagonalization of the $4\times4$ Dirac Hamiltonian. 

\subsection{$t/U$ expansion for the Hubbard model}\label{sec:tU_expansion}

Here we revisit the results of Ref.~\cite{macdonald1988}, which showed how to obtain the Heisenberg effective Hamiltonian from the Hubbard model using the Schrieffer-Wolff transformation. 
The system consists of a lattice with sites described by fermionic operators $c_{i\sigma}$, where $i$ is the lattice site and $\sigma = \uparrow\downarrow$ is the spin. 
The total Hamiltonian is taken to be that of the Fermi-Hubbard model
\begin{equation}\label{tU_H_original}
    H = -t \sum_{\langle i,j\rangle}  \sum_\sigma  c_{i\sigma}^\dagger c_{j\sigma}^{} + U \sum_i n_{i\uparrow} n_{i\downarrow},
\end{equation}
where $t$ is the hopping strength, $U$ is the Coulomb repulsion and  $\langle i,j\rangle$ represents a sum over nearest neighbors in the lattice.

Our goal is to obtain an effective Hamiltonian in the limit where the Coulomb repulsion is strong; i.e., where $t/U \ll 1$. 
In this regime having any site doubly occupied incurs a high energy cost. 
The system will therefore have a tendency to stay in configurations where all sites are singly occupied. 
We can therefore treat the hopping as a perturbation. 
To do that we define 
\begin{equation}
\begin{aligned}
    H_0 &= U \sum_i n_{i\uparrow} n_{i\downarrow}
    \\[0.2cm]
    V &= -t \sum_{\langle i,j\rangle}  \sum_\sigma  \big(c_{i\sigma}^\dagger c_{j\sigma}^{} + c_{j\sigma}^\dagger c_{i\sigma}^{}\big).
\end{aligned}
\end{equation}
To decompose this into eigenoperators we use a trick similar to that in Eq.~\eqref{fermion_decomposition_ci}, and write 
\begin{equation}
    c_{i\sigma}^\dagger c_{j\sigma}^{} = (n_{i\bar{\sigma}}^{} + h_{i\bar{\sigma}}^{})c_{i\sigma}^\dagger c_{j\sigma}^{} (n_{j\bar{\sigma}}^{} + h_{j\bar{\sigma}}^{}),
\end{equation} 
where $\bar{\sigma} = -\sigma$ and $h_{i\sigma} = 1-n_{i\sigma}$. 
We can then split $V$ as
\begin{equation}
    \begin{aligned}
        V_0 &= -t \sum_{\langle i,j\rangle}  \sum_\sigma  \Big( n_{i\bar{\sigma}}^{}c_{i\sigma}^\dagger c_{j\sigma}^{} n_{j\bar{\sigma}}^{}
        +
        h_{i\bar{\sigma}}^{}c_{i\sigma}^\dagger c_{j\sigma}^{} h_{j\bar{\sigma}}^{}
        \Big) + \text{h.c.},
        \\[0.2cm]
        V_+ &= -t\sum_{\langle i,j\rangle}  \sum_\sigma \big(n_{i\bar{\sigma}}^{}c_{i\sigma}^\dagger c_{j\sigma}^{} h_{j\bar{\sigma}}^{} 
        + n_{j\bar{\sigma}}^{} c_{j\sigma}^\dagger c_{i\sigma}^{} h_{i\bar{\sigma}}^{}
        \big),
        \\[0.2cm]
        V_- &= -t\sum_{\langle i,j\rangle}  \sum_\sigma 
        \big(h_{i\bar{\sigma}}^{}c_{i\sigma}^\dagger c_{j\sigma}^{} n_{j\bar{\sigma}}^{}
        + 
        h_{j\bar{\sigma}}^{} c_{j\sigma}^\dagger c_{i\sigma}^{} n_{i\bar{\sigma}}
        \big).
    \end{aligned}
\end{equation}
One may now verify that $V_- = V_+^\dagger$, and   
\begin{equation}
    [H_0,V_0] = 0, \qquad 
    [H_0,V_\pm] = \pm U V_\pm,
\end{equation}
so that $V_0, V_-$ and $V_+$ are all eigenoperators of $H_0$ with frequencies $0,U,-U$ respective.
Since we are assuming $U$ is large, we can therefore associate $V_0=V_d$ and $V_++V_- = V_o$, and treat the latter perturbatively. 
Eq.~\eqref{S_sol} then becomes 
\begin{equation}
    S = -\frac{1}{U}(V_--V_+),
\end{equation}
and Eq.~\eqref{Hprime_eigen_final_d} gives us 
\begin{equation}\label{tU_eigenop_H_SW}
    H' = H_0 + V_0 - \frac{1}{U}[V_-,V_+].
\end{equation}

To proceed, we now need to open up the commutator $[V_-,V_+]$. 
First, notice that we can write $V_- = \sum_{\langle i,j\rangle} V_{-,ij}$, and similarly for $V_+$. The commutators will then be non-zero only if each term has one or both sites in common: 
\begin{equation}
    [V_-,V_+] = \sum_{\langle i,j\rangle} [V_{-,ij}, V_{+,ij}] + \sum_{\langle i,j,k\rangle} [V_{-,ij}, V_{+,jk}],
\end{equation}
where $\langle i,j,k\rangle$ is a sum over triples where $j$ is a nearest neighbor of both $i$ and $k$.     
The leading order contribution turns out to be the first term. For a discussion on the role of the second term, see~\cite{fazekas}. 
Our remaining task is therefore to compute $[V_{-,ij}, V_{+,ij}]$.
This turn out to lead to a lengthy algebra, which goes beyond the scope here, which was to point out the eigenoperator nature of Eq.~\eqref{tU_eigenop_H_SW}. The final result will be the so-called t-J model which, under half-filling, reduces further to the Heisenberg model. 
The interested reader may consult chapter 5 of~\cite{fazekas}.

\section{Discussion and conclusions}

This paper is an attempt to provide a more systematic language for performing time-independent perturbation theory. 
The Schrieffer-Wolff transformation casts the perturbative expansion in the intuitive and modern language of effective Hamiltonians. 
However, actually applying the expansion has always required a certain degree of reverse engineering to figure out the operators that solve Eq.~\eqref{S_equation}. 
The purpose of this paper was to place the spotlight on the semi-obscure method of eigenoperator decomposition, which has so far been restricted mostly to the open quantum systems community. 
As we showed, once the eigenoperator decomposition is known, the application of the method is straightforward. 
And we also discussed more intuitive techniques for finding the eigenoperator decomposition in various systems. 

The design of effective Hamiltonians is a central topic in modern quantum information sciences. 
And it is the hope that this paper simplifies this task. 
One natural example is the development of effective 3-body interactions starting from 2-body terms. 
And natural follow ups include extending the theory to open quantum systems~\cite{kessler2012,sciolla2015,jager2022,malekakhlagh2022,vanhoecke2024}, or to time-periodic Floquet problems~\cite{bukov2016,wang2024a}.

\emph{Acknowledgments - } The author acknowledges extremely fruitful discussions with Ray Parker, Mark Mitchison, Ceren Dag, Federico Rocatti, Eric Andrade, Andr\'e Vieira, Luis Greg\'orio Dias, Annie Schwarz, Chaitanya Murthy. 

\appendix

\section{The Schrieffer-Wolff expansion}\label{app:SW_expansion}

Let us write $H = H_0 + \lambda V$, for a bookkeeping parameter $\lambda$. We expand $S = \lambda S_1 + \lambda^2 S_2 + \ldots$. 
Using the Baker-Campbell-Hausdorff formula and recalling that $V = V_d + V_o$ we get 
\begin{equation}\label{app1_39817981}
    \begin{aligned}
        H' =& e^{S} H e^{-S}
        \\[0.2cm]
        =& H_0 + \lambda\Big(V_d + V_o + [S_1,H_0]\Big) 
        \\[0.2cm]
        &+ \lambda^2 \bigg([S_2,H_0] + [S_1,V_d+V_o] + \tfrac{1}{2}[S_1,[S_1,H_0]]\bigg) 
        \\[0.2cm]
        &+ \lambda^3 \bigg([S_3,H_0] + [S_2,V_d+V_o]
        + \tfrac{1}{2}[S_1,[S_2,H_0]] 
        \\[0.2cm]
        &\quad + \tfrac{1}{2}[S_2,[S_1,H_0]] + \tfrac{1}{2}[S_1,[S_1,V_o+V_d]]
        \\[0.2cm]
        &\quad\quad + \tfrac{1}{3!}[S_1,[S_1,[S_1,H_0]]]
        \bigg)
        + \mathcal{O}(\lambda)^4.
    \end{aligned}
\end{equation}
The basic goal of the SW transformation is to eliminate the block off-diagonal component of order $\lambda$. We must therefore choose 
\begin{equation}
    [H_0,S_1] = V_o,
\end{equation}
Plugging this back in Eq.~\eqref{app1_39817981} leads to 
\begin{equation}\label{app1_98719871987}
    \begin{aligned}
        &H' = H_0 + \lambda V_d + \lambda^2 \bigg( [S_2,H_0] + [S_1,V_d]+ \tfrac{1}{2}[S_1,V_o]\bigg)
        \\[0.2cm]
        &+\lambda^3\Bigg([S_3,H_0] + 
        [S_2,V_d]+\tfrac{1}{2}[S_2,V_o] + \tfrac{1}{2}[S_1,[S_2,H_0]]
        \\[0.2cm]
        &\qquad  + \tfrac{1}{2}[S_1,[S_1,V_d]] + \tfrac{1}{3}[S_1,[S_1,V_o]]
        \Bigg) + \mathcal{O}(\lambda)^4.
    \end{aligned}
\end{equation}
From here onward there are two choices, which lead to distinct types of Schrieffer-Wolff approximations.

First, suppose we perform a basis transformation which involves only a linear in $\lambda$ generator, $S = \lambda S_1$ (i.e., we set $S_2 = S_3 = \ldots = 0$). 
Eq.~\eqref{app1_98719871987} simplifies to 
\begin{equation}\label{app1_SW_case1}
\begin{aligned}
    H' =& H_0 + \lambda V_d + \lambda^2 [S_1,V_d] + \frac{\lambda^2}{2}[S_1,V_o] 
    \\[0.2cm]
    &+ \frac{\lambda^3}{2}[S_1,[S_1,V_d]] + \frac{2\lambda^3}{3!} [S_1,[S_1,V_o]] + \\[0.2cm]
    &+ \frac{\lambda^4}{3!}[S_1,[S_1,[S_1,V_d]]]+\frac{3\lambda^4}{4!}[S_1,[S_1,[S_1,V_o]]] + \ldots,
\end{aligned}    
\end{equation}
and so on.
This choice eliminates the linear term in $V_o$, but does not necessarily lead to a block diagonal Hamiltonian, even at second order. The reason, as discussed in Sec.~\ref{sec:SW}, is that $[S_1,V_o]$ is not necessarily block diagonal. 
However, this kind of transformation has the advantage that we can write an effective Hamiltonian to arbitrarily high orders in $\lambda$, which sometimes still showcases interesting properties. 
The terms containing $V_d$ in Eq.~\eqref{app1_SW_case1} have coefficients of order $\lambda^n$ with prefactors $1/(n-1)!$, while the terms with $V_o$ have $\lambda^n$ multiplied by $(n-1)/n!$.

The second option concerning Eq.~\eqref{app1_98719871987} is to enforce that all terms in the expansion should be block diagonal. 
To this end, $S_1$ alone does not suffice and we need to also introduce the higher order corrections $S_2, S_3,\ldots$. 
Here we proceed up to $\lambda^2$.
Looking at the corresponding term in Eq.~\eqref{app1_98719871987}, 
$[S_1,V_d]$ is strictly block off-diagonal but $[S_1,V_o]$ might have diagonal and off-diagonal components. 
To eliminate all off-diagonals we must therefore choose 
\begin{equation}
    [H_0,S_2] = [S_1,V_d] + \tfrac{1}{2}[S_1,V_o]_o,
\end{equation}
which then leads to 
\begin{equation}
\begin{aligned}
    H' =& H_0 + \lambda V_d + \tfrac{\lambda^2}{2}[S_1,V_o]_d ,
\end{aligned}    
\end{equation}
which is the same result in Eq.~\eqref{Hprime_final_not_d} in the main text.

\section{Schrieffer-Wolff's original calculation}\label{sec:SW_original}

In this section we reproduce the results of~\cite{schrieffer1966} using the new language of eigenoperators developed in Sec.~\ref{sec:main}. 
The original model is the Anderson Hamiltonian describing the interaction of conduction electrons, with annihilation operators $c_{ks}$ (where $s=\pm1$ is the spin), interacting with a single  impurity orbital described by an annihilation operator $c_{ds}$. The unperturbed Hamiltonian $H_0$, describing the two parts independently, is given by 
\begin{equation}\label{Anderson_H0}
    H_0 = \sum_{k,s} \epsilon_k n_{ks} + \sum_s \epsilon_d n_{ds} + U n_{d\uparrow} n_{d\downarrow},
\end{equation}
where $n_{ks} = c_{ks}^\dagger c_{ks}^{}$ and $n_{ds} = c_{ds}^\dagger c_{ds}^{}$.
Here $\epsilon_k$ and $\epsilon_d$ are the single-particle energies of the conduction electrons and the impurity. 
The third term is the Fermi-Hubbard electrostatic repulsion between two opposite spin electrons in  the impurity orbital. 
The interaction between the conduction electrons and the impurity, on the other hand, is given by 
\begin{equation}\label{Anderson_V}
    V = \sum_{k,s} \big\{ V_{kd}^{} c_{ks}^\dagger c_{ds}^{} + V_{kd}^* c_{ds}^\dagger c_{ks}^{}\big\},
\end{equation}
with generic coefficients $V_{kd}$. 

To find the eigenoperator decomposition, we proceed as in Sec.~\ref{sec:eigen_ops_fermi_Hubbard}, particularly Eq.~\eqref{fermion_decomposition_ci}. 
Let $h_{ds} = 1- n_{ds}$ denote the hole operator for the impurity. 
Using that $n_{ds} + h_{ds} = 1$ we can then write Eq.~\eqref{Anderson_V} as
\begin{equation}
    V = \sum_{k,s} \Big\{ V_{kd} \Big(c_{ks}^\dagger c_{ds}^{} n_{d,-s} 
    + c_{ks}^\dagger c_{ds}^{} h_{d,-s}\Big) 
    + V_{kd}^* \Big(
    c_{ds}^\dagger c_{ks}^{} n_{ds} 
    + c_{ds}^\dagger c_{ks}^{} h_{ds} 
    \Big)\Big\}.
\end{equation}
Using the same steps delineated in Sec.~\ref{sec:eigen_ops_fermi_Hubbard}, one can now readily check that each term in this sum is an eigenoperator  of $H_0$. 
That is, the eigenoperators $V_\omega$ correspond to the set
\begin{equation}\label{anderson_eigenoperators}
    \begin{aligned}
        V_{\epsilon_d+U-\epsilon_k} &= \sum_s V_{kd} c_{ks}^\dagger c_{ds}^{} n_{d,-s}^{},
        \\[0.2cm]
        V_{\epsilon_d -\epsilon_k} &= \sum_s V_{kd} c_{ks}^\dagger c_{ds}^{} h_{d,-s},
    \end{aligned}
\end{equation}
as well as their adjoints. 
All elements of $V$ are assumed to belong to $V_o$.
Plugging~\eqref{anderson_eigenoperators} in Eq.~\eqref{S_sol} then yields 
\begin{equation}
    S = - \sum_{k,s} \Bigg\{
    \frac{V_{kd}}{\epsilon_d + U - \epsilon_k} 
    c_{ks}^\dagger c_{ds}^{} n_{d,-s}^{}+ 
    \frac{V_{kd}}{\epsilon_d - \epsilon_k} 
    c_{ks}^\dagger c_{ds}^{} h_{d,-s}^{}\Bigg\} - {\text{h.c.}},
\end{equation}
which is the same as Eq.~(6) of~\cite{schrieffer1966}.

In Ref.~\cite{schrieffer1966} the transformed Hamiltonian was computed by plugging $S$ in Eq.~\eqref{Hprime_final_not_d}. 
Instead, we can now simplify the procedure and jump directly to 
Eq.~\eqref{Hprime_eigen_final_d}.
We will require the following commutation relations: 
\begin{equation}
    \begin{aligned}
        [V_{\epsilon_d + U - \epsilon_k},V_{\epsilon_d + U - \epsilon_q}^\dagger] =& V_{kd}V_{qd}^* \sum_s c_{ks}^\dagger c_{qs}^{} n_{d,-s} + c_{ks}^\dagger c_{q,-s}^{} c_{ds}^{} c_{d,-s}^\dagger
        \\[0.2cm]
        &-\delta_{k,q}|V_{kd}|^2 \sum_s n_{ds} n_{d,-s} ,
        \\[0.3cm]        
        [V_{\epsilon_d  - \epsilon_k},V_{\epsilon_d  - \epsilon_q}^\dagger] =&
        V_{kd}V_{qd}^* \sum_s c_{ks}^\dagger c_{qs}^{} h_{d,-s} - c_{ks}^\dagger c_{q,-s}^{} c_{ds}^{} c_{d,-s}^\dagger
        \\[0.2cm]
        &-\delta_{k,q}|V_{kd}|^2 \sum_s n_{ds} h_{d,-s} ,
        \\[0.3cm]
        [V_{\epsilon_d  +U - \epsilon_k},V_{\epsilon_d  - \epsilon_q}^\dagger] =&
        [V_{\epsilon_d  - \epsilon_k},V_{\epsilon_d +U  - \epsilon_q}^\dagger] = 0.
    \end{aligned}
\end{equation}
The two commutators which are not zero are both associated with a net energy difference $\omega -\omega' = \epsilon_k-\epsilon_q$. 
They therefore are not fully diagonal, but only block diagonal in the energy bands of the conduction electrons. 

Plugging these in Eq.~\eqref{Hprime_eigen_final_d} and organizing the different terms yields 
\begin{equation}\label{anderson_H_final_0}
\begin{aligned}
    H' =& H_0 - \left(\sum_k W_{kk} \right) \sum_s n_{ds} - \left(\tfrac{1}{2}\sum_k J_{kk} \right) n_{d\uparrow} n_{d\downarrow}
    \\[0.2cm]
    &+ \sum_{k,q,s} \left(W_{kq} + \tfrac{J_{kq}}{2}  n_{d,-s} \right) c_{ks}^\dagger c_{qs}^{}
    - \sum_{k,q,s} \tfrac{J_{kq}}{2} c_{ks}^\dagger c_{qs}^{} c_{d,-s}^\dagger c_{ds},
\end{aligned}
\end{equation}
where 
\begin{align}\label{Anderson_Wkq}
    W_{kq} &=  \frac{V_{kd}V_{qd}^*}{2}\left(\frac{1}{\epsilon_k-\epsilon_d}+\frac{1}{\epsilon_q-\epsilon_d} \right),
    \\[0.2cm]\label{Anderson_Jkq}
    J_{kq} &= V_{kd}V_{qd}^* \left( 
    \frac{1}{\epsilon_k - \epsilon_d - U}+
    \frac{1}{\epsilon_q - \epsilon_d - U}
    \right.
    \\[0.2cm]\nonumber
    &\qquad\qquad\qquad\quad \left. 
    -\frac{1}{\epsilon_k - \epsilon_d}
    -\frac{1}{\epsilon_q - \epsilon_d}
    \right).
\end{align}
The two terms in the first line of~\eqref{anderson_H_final_0} act as renormalizations of $\epsilon_d$ and $U$ in Eq.~\eqref{Anderson_H0}. 
In the second line, the first term is a impurity-dependent scattering of the conduction electrons, which scatter from $q\to k$ with amplitude given either by $W_{kq}$, when $n_{d,-s}=0$, or $W_{kq}+J_{kq}$, when $n_{d,-s}=1$. 
Finally, the last term is an exchange interaction in the Kondo model, which is precisely the effect that Schrieffer and Wolff intended to capture.  

To make the latter more evident, and to better connect Eq.~\eqref{anderson_H_final_0} with the results in~\cite{schrieffer1966}, we  introduce two component spinors 
\begin{equation}
    \Psi_k = \begin{pmatrix}
        c_{k\uparrow} \\ c_{k\downarrow}
    \end{pmatrix},
    \qquad 
    \Psi_d = \begin{pmatrix}
        c_{d\uparrow} \\ c_{d\downarrow}
    \end{pmatrix},
\end{equation}
and use the identity 
\begin{equation}
\begin{aligned}
    \sum_s c_{ks}^\dagger c_{q,-s}^{} c_{d,-s}^\dagger c_{ds} =& \tfrac{1}{2}(\Psi_k^\dagger \bm{\sigma}\Psi_q^{}) (\Psi_d^\dagger \bm{\sigma}\Psi_d^{})
    \\[0.2cm]
    \qquad \qquad &- \tfrac{1}{2}\sum_s c_{ks}^\dagger c_{qs}^{} (n_{ds} - n_{d,-s}),
\end{aligned}    
\end{equation}
where $\bm{\sigma} = (\sigma_x,\sigma_y,\sigma_z)$ are the Pauli matrices.
As a result, we find after some rearranging:
\begin{equation}
    \begin{aligned}
        H' &= H_0 - \left(\sum_k W_{kk} \right) \sum_s n_{ds} - \left(\tfrac{1}{2}\sum_k J_{kk} \right) n_{d\uparrow} n_{d\downarrow}
    \\[0.2cm]
    &+\sum_{k,q} \left( W_{kq} + \frac{J_{kq}}{4} \Psi_d^\dagger \Psi_d^{}\right) \Psi_k^\dagger \Psi_q
    \\[0.2cm]
    &- \sum_{k,q} \frac{J_{kq}}{4}(\Psi_k^\dagger \bm{\sigma} \Psi_q^{}) (\Psi_d^\dagger \bm{\sigma} \Psi_d^{}).
    \end{aligned}
\end{equation}
The last term is exactly the exchange interaction.

It is worth emphasizing that the main calculations in this section essentially ended in Eq.~\eqref{anderson_eigenoperators}, with the identification of the eigenoperators. 
Everything else was merely a (admittedly cumbersome) rearrangement/simplification of the final Hamiltonian. 
This is made possible by the convenience of Eq.~\eqref{Hprime_eigen_final_d}.

\bibliography{references}

\begin{thebibliography}{45}%
\makeatletter
\providecommand \@ifxundefined [1]{%
 \@ifx{#1\undefined}
}%
\providecommand \@ifnum [1]{%
 \ifnum #1\expandafter \@firstoftwo
 \else \expandafter \@secondoftwo
 \fi
}%
\providecommand \@ifx [1]{%
 \ifx #1\expandafter \@firstoftwo
 \else \expandafter \@secondoftwo
 \fi
}%
\providecommand \natexlab [1]{#1}%
\providecommand \enquote  [1]{``#1''}%
\providecommand \bibnamefont  [1]{#1}%
\providecommand \bibfnamefont [1]{#1}%
\providecommand \citenamefont [1]{#1}%
\providecommand \href@noop [0]{\@secondoftwo}%
\providecommand \href [0]{\begingroup \@sanitize@url \@href}%
\providecommand \@href[1]{\@@startlink{#1}\@@href}%
\providecommand \@@href[1]{\endgroup#1\@@endlink}%
\providecommand \@sanitize@url [0]{\catcode `\\12\catcode `\$12\catcode `\&12\catcode `\#12\catcode `\^12\catcode `\_12\catcode `\%12\relax}%
\providecommand \@@startlink[1]{}%
\providecommand \@@endlink[0]{}%
\providecommand \url  [0]{\begingroup\@sanitize@url \@url }%
\providecommand \@url [1]{\endgroup\@href {#1}{\urlprefix }}%
\providecommand \urlprefix  [0]{URL }%
\providecommand \Eprint [0]{\href }%
\providecommand \doibase [0]{http://dx.doi.org/}%
\providecommand \selectlanguage [0]{\@gobble}%
\providecommand \bibinfo  [0]{\@secondoftwo}%
\providecommand \bibfield  [0]{\@secondoftwo}%
\providecommand \translation [1]{[#1]}%
\providecommand \BibitemOpen [0]{}%
\providecommand \bibitemStop [0]{}%
\providecommand \bibitemNoStop [0]{.\EOS\space}%
\providecommand \EOS [0]{\spacefactor3000\relax}%
\providecommand \BibitemShut  [1]{\csname bibitem#1\endcsname}%
\let\auto@bib@innerbib\@empty
\bibitem [{\citenamefont {Rayleigh}(1969)}]{rayleigh1969}%
  \BibitemOpen
  \bibfield  {author} {\bibinfo {author} {\bibfnamefont {J.~W.~S.}\ \bibnamefont {Rayleigh}},\ }\href@noop {} {{\selectlanguage {eng}\emph {\bibinfo {title} {The theory of sound. {Vol}. 1}}}},\ \bibinfo {edition} {repr. of 2. ed. rev. and enl., 1894}\ ed.,\ \bibinfo {series} {Dover classics of science and mathematics}, Vol.~\bibinfo {volume} {1}\ (\bibinfo  {publisher} {Dover Publ},\ \bibinfo {address} {New York},\ \bibinfo {year} {1969})\BibitemShut {NoStop}%
\bibitem [{\citenamefont {Schrödinger}(1926)}]{schrodinger1926}%
  \BibitemOpen
  \bibfield  {author} {\bibinfo {author} {\bibfnamefont {E.}~\bibnamefont {Schrödinger}},\ }\href {\doibase 10.1002/andp.19263851302} {\bibfield  {journal} {\bibinfo  {journal} {Annalen der Physik}\ }\textbf {\bibinfo {volume} {385}},\ \bibinfo {pages} {437} (\bibinfo {year} {1926})}\BibitemShut {NoStop}%
\bibitem [{\citenamefont {Lee}\ \emph {et~al.}(2022)\citenamefont {Lee}, \citenamefont {Tan}, \citenamefont {Nguyen}, \citenamefont {Budoyo}, \citenamefont {Park}, \citenamefont {Hufnagel}, \citenamefont {Yap}, \citenamefont {Møbjerg}, \citenamefont {Vedral}, \citenamefont {Paterek},\ and\ \citenamefont {Dumke}}]{lee2022}%
  \BibitemOpen
  \bibfield  {author} {\bibinfo {author} {\bibfnamefont {K.~S.}\ \bibnamefont {Lee}}, \bibinfo {author} {\bibfnamefont {Y.~P.}\ \bibnamefont {Tan}}, \bibinfo {author} {\bibfnamefont {L.~H.}\ \bibnamefont {Nguyen}}, \bibinfo {author} {\bibfnamefont {R.~P.}\ \bibnamefont {Budoyo}}, \bibinfo {author} {\bibfnamefont {K.~H.}\ \bibnamefont {Park}}, \bibinfo {author} {\bibfnamefont {C.}~\bibnamefont {Hufnagel}}, \bibinfo {author} {\bibfnamefont {Y.~S.}\ \bibnamefont {Yap}}, \bibinfo {author} {\bibfnamefont {N.}~\bibnamefont {Møbjerg}}, \bibinfo {author} {\bibfnamefont {V.}~\bibnamefont {Vedral}}, \bibinfo {author} {\bibfnamefont {T.}~\bibnamefont {Paterek}}, \ and\ \bibinfo {author} {\bibfnamefont {R.}~\bibnamefont {Dumke}},\ }\href {\doibase 10.1088/1367-2630/aca81f} {\bibfield  {journal} {\bibinfo  {journal} {New Journal of Physics}\ }\textbf {\bibinfo {volume} {24}},\ \bibinfo {pages} {123024} (\bibinfo {year} {2022})},\ \bibinfo {note} {publisher: IOP Publishing}\BibitemShut {NoStop}%
\bibitem [{\citenamefont {Schrieffer}\ and\ \citenamefont {Wolff}(1966)}]{schrieffer1966}%
  \BibitemOpen
  \bibfield  {author} {\bibinfo {author} {\bibfnamefont {J.~R.}\ \bibnamefont {Schrieffer}}\ and\ \bibinfo {author} {\bibfnamefont {P.~A.}\ \bibnamefont {Wolff}},\ }\href {\doibase 10.1103/PhysRev.149.491} {\bibfield  {journal} {\bibinfo  {journal} {Physical Review}\ }\textbf {\bibinfo {volume} {149}},\ \bibinfo {pages} {491} (\bibinfo {year} {1966})},\ \bibinfo {note} {publisher: American Physical Society}\BibitemShut {NoStop}%
\bibitem [{\citenamefont {Van~Vleck}(1929)}]{vanvleck1929}%
  \BibitemOpen
  \bibfield  {author} {\bibinfo {author} {\bibfnamefont {J.~H.}\ \bibnamefont {Van~Vleck}},\ }\href {\doibase 10.1103/PhysRev.33.467} {\bibfield  {journal} {\bibinfo  {journal} {Physical Review}\ }\textbf {\bibinfo {volume} {33}},\ \bibinfo {pages} {467} (\bibinfo {year} {1929})}\BibitemShut {NoStop}%
\bibitem [{\citenamefont {Löwdin}(1962)}]{lowdin1962}%
  \BibitemOpen
  \bibfield  {author} {\bibinfo {author} {\bibfnamefont {P.-O.}\ \bibnamefont {Löwdin}},\ }\href {\doibase 10.1063/1.1724312} {\bibfield  {journal} {\bibinfo  {journal} {Journal of Mathematical Physics}\ }\textbf {\bibinfo {volume} {3}},\ \bibinfo {pages} {969} (\bibinfo {year} {1962})}\BibitemShut {NoStop}%
\bibitem [{\citenamefont {Fröhlich}(1952)}]{frohlich1952}%
  \BibitemOpen
  \bibfield  {author} {\bibinfo {author} {\bibfnamefont {H.}~\bibnamefont {Fröhlich}},\ }\href {\doibase 10.1098/rspa.1952.0212} {\bibfield  {journal} {\bibinfo  {journal} {Proceedings of the Royal Society of London. Series A. Mathematical and Physical Sciences}\ }\textbf {\bibinfo {volume} {215}},\ \bibinfo {pages} {291} (\bibinfo {year} {1952})},\ \bibinfo {note} {publisher: Royal Society}\BibitemShut {NoStop}%
\bibitem [{\citenamefont {Foldy}\ and\ \citenamefont {Wouthuysen}(1950)}]{foldy1950}%
  \BibitemOpen
  \bibfield  {author} {\bibinfo {author} {\bibfnamefont {L.~L.}\ \bibnamefont {Foldy}}\ and\ \bibinfo {author} {\bibfnamefont {S.~A.}\ \bibnamefont {Wouthuysen}},\ }\href {\doibase 10.1103/PhysRev.78.29} {\bibfield  {journal} {\bibinfo  {journal} {Physical Review}\ }\textbf {\bibinfo {volume} {78}},\ \bibinfo {pages} {29} (\bibinfo {year} {1950})}\BibitemShut {NoStop}%
\bibitem [{\citenamefont {Bravyi}\ \emph {et~al.}(2011)\citenamefont {Bravyi}, \citenamefont {DiVincenzo},\ and\ \citenamefont {Loss}}]{bravyi2011}%
  \BibitemOpen
  \bibfield  {author} {\bibinfo {author} {\bibfnamefont {S.}~\bibnamefont {Bravyi}}, \bibinfo {author} {\bibfnamefont {D.~P.}\ \bibnamefont {DiVincenzo}}, \ and\ \bibinfo {author} {\bibfnamefont {D.}~\bibnamefont {Loss}},\ }\href {\doibase 10.1016/j.aop.2011.06.004} {\bibfield  {journal} {\bibinfo  {journal} {Annals of Physics}\ }\textbf {\bibinfo {volume} {326}},\ \bibinfo {pages} {2793} (\bibinfo {year} {2011})},\ \bibinfo {note} {arXiv: 1105.0675}\BibitemShut {NoStop}%
\bibitem [{\citenamefont {Datta}\ \emph {et~al.}(2011)\citenamefont {Datta}, \citenamefont {Frohlich},\ and\ \citenamefont {Rey-Bellet}}]{datta2011}%
  \BibitemOpen
  \bibfield  {author} {\bibinfo {author} {\bibfnamefont {N.}~\bibnamefont {Datta}}, \bibinfo {author} {\bibfnamefont {J.}~\bibnamefont {Frohlich}}, \ and\ \bibinfo {author} {\bibfnamefont {L.}~\bibnamefont {Rey-Bellet}},\ }\href@noop {} {\  (\bibinfo {year} {2011})}\BibitemShut {NoStop}%
\bibitem [{\citenamefont {Bukov}\ \emph {et~al.}(2016)\citenamefont {Bukov}, \citenamefont {Kolodrubetz},\ and\ \citenamefont {Polkovnikov}}]{bukov2016}%
  \BibitemOpen
  \bibfield  {author} {\bibinfo {author} {\bibfnamefont {M.}~\bibnamefont {Bukov}}, \bibinfo {author} {\bibfnamefont {M.}~\bibnamefont {Kolodrubetz}}, \ and\ \bibinfo {author} {\bibfnamefont {A.}~\bibnamefont {Polkovnikov}},\ }\href {\doibase 10.1103/PhysRevLett.116.125301} {\bibfield  {journal} {\bibinfo  {journal} {Physical Review Letters}\ }\textbf {\bibinfo {volume} {116}},\ \bibinfo {pages} {125301} (\bibinfo {year} {2016})},\ \bibinfo {note} {publisher: American Physical Society}\BibitemShut {NoStop}%
\bibitem [{\citenamefont {Haq}\ and\ \citenamefont {Singh}(2020)}]{haq2020}%
  \BibitemOpen
  \bibfield  {author} {\bibinfo {author} {\bibfnamefont {R.~U.}\ \bibnamefont {Haq}}\ and\ \bibinfo {author} {\bibfnamefont {K.}~\bibnamefont {Singh}},\ }\href {http://arxiv.org/abs/2004.06534} {{\selectlanguage {en}\enquote {\bibinfo {title} {A systematic method for {Schrieffer}-{Wolff} transformation and its generalizations},}\ }} (\bibinfo {year} {2020}),\ \bibinfo {note} {arXiv:2004.06534 [cond-mat]}\BibitemShut {NoStop}%
\bibitem [{\citenamefont {Kale}\ \emph {et~al.}(2022)\citenamefont {Kale}, \citenamefont {Huhn}, \citenamefont {Xu}, \citenamefont {Kendrick}, \citenamefont {Lebrat}, \citenamefont {Chiu}, \citenamefont {Ji}, \citenamefont {Grusdt}, \citenamefont {Bohrdt},\ and\ \citenamefont {Greiner}}]{kale2022}%
  \BibitemOpen
  \bibfield  {author} {\bibinfo {author} {\bibfnamefont {A.}~\bibnamefont {Kale}}, \bibinfo {author} {\bibfnamefont {J.~H.}\ \bibnamefont {Huhn}}, \bibinfo {author} {\bibfnamefont {M.}~\bibnamefont {Xu}}, \bibinfo {author} {\bibfnamefont {L.~H.}\ \bibnamefont {Kendrick}}, \bibinfo {author} {\bibfnamefont {M.}~\bibnamefont {Lebrat}}, \bibinfo {author} {\bibfnamefont {C.}~\bibnamefont {Chiu}}, \bibinfo {author} {\bibfnamefont {G.}~\bibnamefont {Ji}}, \bibinfo {author} {\bibfnamefont {F.}~\bibnamefont {Grusdt}}, \bibinfo {author} {\bibfnamefont {A.}~\bibnamefont {Bohrdt}}, \ and\ \bibinfo {author} {\bibfnamefont {M.}~\bibnamefont {Greiner}},\ }\href {\doibase 10.1103/PhysRevA.106.012428} {\bibfield  {journal} {\bibinfo  {journal} {Physical Review A}\ }\textbf {\bibinfo {volume} {106}},\ \bibinfo {pages} {012428} (\bibinfo {year} {2022})},\ \bibinfo {note} {arXiv:2203.07366 [cond-mat, physics:quant-ph]}\BibitemShut {NoStop}%
\bibitem [{\citenamefont {Day}\ \emph {et~al.}(2024)\citenamefont {Day}, \citenamefont {Miles}, \citenamefont {Kerstens}, \citenamefont {Varjas},\ and\ \citenamefont {Akhmerov}}]{day2024}%
  \BibitemOpen
  \bibfield  {author} {\bibinfo {author} {\bibfnamefont {I.~A.}\ \bibnamefont {Day}}, \bibinfo {author} {\bibfnamefont {S.}~\bibnamefont {Miles}}, \bibinfo {author} {\bibfnamefont {H.~K.}\ \bibnamefont {Kerstens}}, \bibinfo {author} {\bibfnamefont {D.}~\bibnamefont {Varjas}}, \ and\ \bibinfo {author} {\bibfnamefont {A.~R.}\ \bibnamefont {Akhmerov}},\ }\href {http://arxiv.org/abs/2404.03728} {{\selectlanguage {en}\enquote {\bibinfo {title} {Pymablock: an algorithm and a package for quasi-degenerate perturbation theory},}\ }} (\bibinfo {year} {2024}),\ \bibinfo {note} {arXiv:2404.03728 [cond-mat, physics:quant-ph]}\BibitemShut {NoStop}%
\bibitem [{\citenamefont {Dag}\ and\ \citenamefont {Rokaj}(2023)}]{dag2023}%
  \BibitemOpen
  \bibfield  {author} {\bibinfo {author} {\bibfnamefont {C.~B.}\ \bibnamefont {Dag}}\ and\ \bibinfo {author} {\bibfnamefont {V.}~\bibnamefont {Rokaj}},\ }\href {http://arxiv.org/abs/2311.02806} {\  (\bibinfo {year} {2023})},\ \bibinfo {note} {arXiv:2311.02806 [cond-mat]}\BibitemShut {NoStop}%
\bibitem [{\citenamefont {Hu}\ \emph {et~al.}(2023)\citenamefont {Hu}, \citenamefont {Bernevig},\ and\ \citenamefont {Tsvelik}}]{hu2023}%
  \BibitemOpen
  \bibfield  {author} {\bibinfo {author} {\bibfnamefont {H.}~\bibnamefont {Hu}}, \bibinfo {author} {\bibfnamefont {B.~A.}\ \bibnamefont {Bernevig}}, \ and\ \bibinfo {author} {\bibfnamefont {A.~M.}\ \bibnamefont {Tsvelik}},\ }\href {\doibase 10.1103/PhysRevLett.131.026502} {\bibfield  {journal} {\bibinfo  {journal} {Physical Review Letters}\ }\textbf {\bibinfo {volume} {131}},\ \bibinfo {pages} {026502} (\bibinfo {year} {2023})},\ \bibinfo {note} {arXiv:2301.04669 [cond-mat]}\BibitemShut {NoStop}%
\bibitem [{\citenamefont {Cohen-Tannoudji}\ \emph {et~al.}(1998)\citenamefont {Cohen-Tannoudji}, \citenamefont {Dupont-Roc},\ and\ \citenamefont {Grynberg}}]{CohenTannoudji1998}%
  \BibitemOpen
  \bibfield  {author} {\bibinfo {author} {\bibfnamefont {C.}~\bibnamefont {Cohen-Tannoudji}}, \bibinfo {author} {\bibfnamefont {J.}~\bibnamefont {Dupont-Roc}}, \ and\ \bibinfo {author} {\bibfnamefont {G.}~\bibnamefont {Grynberg}},\ }\href {\doibase 10.1002/9783527617197.fmatter} {{\selectlanguage {en}\emph {\bibinfo {title} {Atom—{Photon} {Interactions}: {Basic} {Process} and {Applications}}}}}\ (\bibinfo  {publisher} {John Wiley \& Sons, Ltd},\ \bibinfo {year} {1998})\ \bibinfo {note} {\_eprint: https://onlinelibrary.wiley.com/doi/pdf/10.1002/9783527617197.fmatter}\BibitemShut {NoStop}%
\bibitem [{\citenamefont {Gerry}\ and\ \citenamefont {Knight}()}]{gerry}%
  \BibitemOpen
  \bibfield  {author} {\bibinfo {author} {\bibfnamefont {C.~C.}\ \bibnamefont {Gerry}}\ and\ \bibinfo {author} {\bibfnamefont {P.~L.}\ \bibnamefont {Knight}},\ }\href@noop {} {\emph {\bibinfo {title} {Introductory {Quantum} {Optics}}}}\ (\bibinfo  {publisher} {Cambridge University Press})\BibitemShut {NoStop}%
\bibitem [{\citenamefont {Krantz}\ \emph {et~al.}(2019)\citenamefont {Krantz}, \citenamefont {Kjaergaard}, \citenamefont {Yan}, \citenamefont {Orlando}, \citenamefont {Gustavsson},\ and\ \citenamefont {Oliver}}]{krantz2019}%
  \BibitemOpen
  \bibfield  {author} {\bibinfo {author} {\bibfnamefont {P.}~\bibnamefont {Krantz}}, \bibinfo {author} {\bibfnamefont {M.}~\bibnamefont {Kjaergaard}}, \bibinfo {author} {\bibfnamefont {F.}~\bibnamefont {Yan}}, \bibinfo {author} {\bibfnamefont {T.~P.}\ \bibnamefont {Orlando}}, \bibinfo {author} {\bibfnamefont {S.}~\bibnamefont {Gustavsson}}, \ and\ \bibinfo {author} {\bibfnamefont {W.~D.}\ \bibnamefont {Oliver}},\ }\href {\doibase 10.1063/1.5089550} {\bibfield  {journal} {\bibinfo  {journal} {Applied Physics Reviews}\ }\textbf {\bibinfo {volume} {6}},\ \bibinfo {pages} {021318} (\bibinfo {year} {2019})}\BibitemShut {NoStop}%
\bibitem [{\citenamefont {Mercurio}\ \emph {et~al.}(2024)\citenamefont {Mercurio}, \citenamefont {Russo}, \citenamefont {Mauceri}, \citenamefont {Savasta}, \citenamefont {Nori}, \citenamefont {Macrì},\ and\ \citenamefont {Franco}}]{mercurio2024}%
  \BibitemOpen
  \bibfield  {author} {\bibinfo {author} {\bibfnamefont {A.}~\bibnamefont {Mercurio}}, \bibinfo {author} {\bibfnamefont {E.}~\bibnamefont {Russo}}, \bibinfo {author} {\bibfnamefont {F.}~\bibnamefont {Mauceri}}, \bibinfo {author} {\bibfnamefont {S.}~\bibnamefont {Savasta}}, \bibinfo {author} {\bibfnamefont {F.}~\bibnamefont {Nori}}, \bibinfo {author} {\bibfnamefont {V.}~\bibnamefont {Macrì}}, \ and\ \bibinfo {author} {\bibfnamefont {R.~L.}\ \bibnamefont {Franco}},\ }\href {\doibase 10.48550/arXiv.2402.04339} {\enquote {\bibinfo {title} {Bilateral photon emission from a vibrating mirror and multiphoton entanglement generation},}\ } (\bibinfo {year} {2024}),\ \bibinfo {note} {arXiv:2402.04339 [quant-ph]}\BibitemShut {NoStop}%
\bibitem [{\citenamefont {Blais}\ \emph {et~al.}(2004)\citenamefont {Blais}, \citenamefont {Huang}, \citenamefont {Wallraff}, \citenamefont {Girvin},\ and\ \citenamefont {Schoelkopf}}]{blais2004}%
  \BibitemOpen
  \bibfield  {author} {\bibinfo {author} {\bibfnamefont {A.}~\bibnamefont {Blais}}, \bibinfo {author} {\bibfnamefont {R.-S.}\ \bibnamefont {Huang}}, \bibinfo {author} {\bibfnamefont {A.}~\bibnamefont {Wallraff}}, \bibinfo {author} {\bibfnamefont {S.~M.}\ \bibnamefont {Girvin}}, \ and\ \bibinfo {author} {\bibfnamefont {R.~J.}\ \bibnamefont {Schoelkopf}},\ }\href {\doibase 10.1103/PhysRevA.69.062320} {\bibfield  {journal} {\bibinfo  {journal} {Physical Review A}\ }\textbf {\bibinfo {volume} {69}},\ \bibinfo {pages} {062320} (\bibinfo {year} {2004})}\BibitemShut {NoStop}%
\bibitem [{\citenamefont {Wallraff}\ \emph {et~al.}(2004)\citenamefont {Wallraff}, \citenamefont {Schuster}, \citenamefont {Blais}, \citenamefont {Frunzio}, \citenamefont {Huang}, \citenamefont {Majer}, \citenamefont {Kumar}, \citenamefont {Girvin},\ and\ \citenamefont {Schoelkopf}}]{wallraff2004}%
  \BibitemOpen
  \bibfield  {author} {\bibinfo {author} {\bibfnamefont {A.}~\bibnamefont {Wallraff}}, \bibinfo {author} {\bibfnamefont {D.~I.}\ \bibnamefont {Schuster}}, \bibinfo {author} {\bibfnamefont {A.}~\bibnamefont {Blais}}, \bibinfo {author} {\bibfnamefont {L.}~\bibnamefont {Frunzio}}, \bibinfo {author} {\bibfnamefont {R.-S.}\ \bibnamefont {Huang}}, \bibinfo {author} {\bibfnamefont {J.}~\bibnamefont {Majer}}, \bibinfo {author} {\bibfnamefont {S.}~\bibnamefont {Kumar}}, \bibinfo {author} {\bibfnamefont {S.~M.}\ \bibnamefont {Girvin}}, \ and\ \bibinfo {author} {\bibfnamefont {R.~J.}\ \bibnamefont {Schoelkopf}},\ }\href {\doibase 10.1038/nature02851} {\bibfield  {journal} {\bibinfo  {journal} {Nature}\ }\textbf {\bibinfo {volume} {431}},\ \bibinfo {pages} {162} (\bibinfo {year} {2004})},\ \bibinfo {note} {publisher: Nature Publishing Group}\BibitemShut {NoStop}%
\bibitem [{\citenamefont {Koch}\ \emph {et~al.}(2007)\citenamefont {Koch}, \citenamefont {Yu}, \citenamefont {Gambetta}, \citenamefont {Houck}, \citenamefont {Schuster}, \citenamefont {Majer}, \citenamefont {Blais}, \citenamefont {Devoret}, \citenamefont {Girvin},\ and\ \citenamefont {Schoelkopf}}]{koch2007}%
  \BibitemOpen
  \bibfield  {author} {\bibinfo {author} {\bibfnamefont {J.}~\bibnamefont {Koch}}, \bibinfo {author} {\bibfnamefont {T.~M.}\ \bibnamefont {Yu}}, \bibinfo {author} {\bibfnamefont {J.}~\bibnamefont {Gambetta}}, \bibinfo {author} {\bibfnamefont {A.~A.}\ \bibnamefont {Houck}}, \bibinfo {author} {\bibfnamefont {D.~I.}\ \bibnamefont {Schuster}}, \bibinfo {author} {\bibfnamefont {J.}~\bibnamefont {Majer}}, \bibinfo {author} {\bibfnamefont {A.}~\bibnamefont {Blais}}, \bibinfo {author} {\bibfnamefont {M.~H.}\ \bibnamefont {Devoret}}, \bibinfo {author} {\bibfnamefont {S.~M.}\ \bibnamefont {Girvin}}, \ and\ \bibinfo {author} {\bibfnamefont {R.~J.}\ \bibnamefont {Schoelkopf}},\ }\href {\doibase 10.1103/PhysRevA.76.042319} {\bibfield  {journal} {\bibinfo  {journal} {Physical Review A}\ }\textbf {\bibinfo {volume} {76}},\ \bibinfo {pages} {042319} (\bibinfo {year} {2007})},\ \bibinfo {note} {arXiv:cond-mat/0703002}\BibitemShut {NoStop}%
\bibitem [{\citenamefont {Boissonneault}\ \emph {et~al.}(2009)\citenamefont {Boissonneault}, \citenamefont {Gambetta},\ and\ \citenamefont {Blais}}]{boissonneault2009}%
  \BibitemOpen
  \bibfield  {author} {\bibinfo {author} {\bibfnamefont {M.}~\bibnamefont {Boissonneault}}, \bibinfo {author} {\bibfnamefont {J.~M.}\ \bibnamefont {Gambetta}}, \ and\ \bibinfo {author} {\bibfnamefont {A.}~\bibnamefont {Blais}},\ }\href {\doibase 10.1103/PhysRevA.79.013819} {\bibfield  {journal} {\bibinfo  {journal} {Physical Review A}\ }\textbf {\bibinfo {volume} {79}},\ \bibinfo {pages} {013819} (\bibinfo {year} {2009})},\ \bibinfo {note} {arXiv:0810.1336 [cond-mat, physics:quant-ph]}\BibitemShut {NoStop}%
\bibitem [{\citenamefont {Blais}\ \emph {et~al.}(2021)\citenamefont {Blais}, \citenamefont {Grimsmo}, \citenamefont {Girvin},\ and\ \citenamefont {Wallraff}}]{blais2021}%
  \BibitemOpen
  \bibfield  {author} {\bibinfo {author} {\bibfnamefont {A.}~\bibnamefont {Blais}}, \bibinfo {author} {\bibfnamefont {A.~L.}\ \bibnamefont {Grimsmo}}, \bibinfo {author} {\bibfnamefont {S.~M.}\ \bibnamefont {Girvin}}, \ and\ \bibinfo {author} {\bibfnamefont {A.}~\bibnamefont {Wallraff}},\ }\href {\doibase 10.1103/RevModPhys.93.025005} {\bibfield  {journal} {\bibinfo  {journal} {Reviews of Modern Physics}\ }\textbf {\bibinfo {volume} {93}},\ \bibinfo {pages} {025005} (\bibinfo {year} {2021})},\ \bibinfo {note} {arXiv:2005.12667 [quant-ph]}\BibitemShut {NoStop}%
\bibitem [{\citenamefont {Kessler}(2012)}]{kessler2012}%
  \BibitemOpen
  \bibfield  {author} {\bibinfo {author} {\bibfnamefont {E.~M.}\ \bibnamefont {Kessler}},\ }\href {\doibase 10.1103/PhysRevA.86.012126} {\bibfield  {journal} {\bibinfo  {journal} {Physical Review A}\ }\textbf {\bibinfo {volume} {86}},\ \bibinfo {pages} {012126} (\bibinfo {year} {2012})}\BibitemShut {NoStop}%
\bibitem [{\citenamefont {Sciolla}\ \emph {et~al.}(2015)\citenamefont {Sciolla}, \citenamefont {Poletti},\ and\ \citenamefont {Kollath}}]{sciolla2015}%
  \BibitemOpen
  \bibfield  {author} {\bibinfo {author} {\bibfnamefont {B.}~\bibnamefont {Sciolla}}, \bibinfo {author} {\bibfnamefont {D.}~\bibnamefont {Poletti}}, \ and\ \bibinfo {author} {\bibfnamefont {C.}~\bibnamefont {Kollath}},\ }\href {\doibase 10.1103/PhysRevLett.114.170401} {\bibfield  {journal} {\bibinfo  {journal} {Physical Review Letters}\ }\textbf {\bibinfo {volume} {114}},\ \bibinfo {pages} {170401} (\bibinfo {year} {2015})},\ \bibinfo {note} {arXiv:1407.4939 [cond-mat, physics:quant-ph]}\BibitemShut {NoStop}%
\bibitem [{\citenamefont {Jäger}\ \emph {et~al.}(2022)\citenamefont {Jäger}, \citenamefont {Schmit}, \citenamefont {Morigi}, \citenamefont {Holland},\ and\ \citenamefont {Betzholz}}]{jager2022}%
  \BibitemOpen
  \bibfield  {author} {\bibinfo {author} {\bibfnamefont {S.~B.}\ \bibnamefont {Jäger}}, \bibinfo {author} {\bibfnamefont {T.}~\bibnamefont {Schmit}}, \bibinfo {author} {\bibfnamefont {G.}~\bibnamefont {Morigi}}, \bibinfo {author} {\bibfnamefont {M.~J.}\ \bibnamefont {Holland}}, \ and\ \bibinfo {author} {\bibfnamefont {R.}~\bibnamefont {Betzholz}},\ }\href {\doibase 10.1103/PhysRevLett.129.063601} {\bibfield  {journal} {\bibinfo  {journal} {Physical Review Letters}\ }\textbf {\bibinfo {volume} {129}},\ \bibinfo {pages} {063601} (\bibinfo {year} {2022})},\ \bibinfo {note} {arXiv:2203.03302 [cond-mat, physics:quant-ph]}\BibitemShut {NoStop}%
\bibitem [{\citenamefont {Malekakhlagh}\ \emph {et~al.}(2022)\citenamefont {Malekakhlagh}, \citenamefont {Magesan},\ and\ \citenamefont {Govia}}]{malekakhlagh2022}%
  \BibitemOpen
  \bibfield  {author} {\bibinfo {author} {\bibfnamefont {M.}~\bibnamefont {Malekakhlagh}}, \bibinfo {author} {\bibfnamefont {E.}~\bibnamefont {Magesan}}, \ and\ \bibinfo {author} {\bibfnamefont {L.~C.~G.}\ \bibnamefont {Govia}},\ }\href {\doibase 10.1103/PhysRevA.106.052601} {\bibfield  {journal} {\bibinfo  {journal} {Physical Review A}\ }\textbf {\bibinfo {volume} {106}},\ \bibinfo {pages} {052601} (\bibinfo {year} {2022})},\ \bibinfo {note} {arXiv:2206.12467 [cond-mat, physics:quant-ph]}\BibitemShut {NoStop}%
\bibitem [{\citenamefont {Vanhoecke}\ and\ \citenamefont {Schirò}(2024)}]{vanhoecke2024}%
  \BibitemOpen
  \bibfield  {author} {\bibinfo {author} {\bibfnamefont {M.}~\bibnamefont {Vanhoecke}}\ and\ \bibinfo {author} {\bibfnamefont {M.}~\bibnamefont {Schirò}},\ }\href {http://arxiv.org/abs/2405.17348} {{\selectlanguage {en}\enquote {\bibinfo {title} {Kondo-{Zeno} crossover in the dynamics of a monitored quantum dot},}\ }} (\bibinfo {year} {2024}),\ \bibinfo {note} {arXiv:2405.17348 [cond-mat, physics:quant-ph]}\BibitemShut {NoStop}%
\bibitem [{\citenamefont {Wang}\ \emph {et~al.}(2024)\citenamefont {Wang}, \citenamefont {Méndez-Córdoba}, \citenamefont {Jaksch},\ and\ \citenamefont {Schlawin}}]{wang2024a}%
  \BibitemOpen
  \bibfield  {author} {\bibinfo {author} {\bibfnamefont {X.}~\bibnamefont {Wang}}, \bibinfo {author} {\bibfnamefont {F.~P.~M.}\ \bibnamefont {Méndez-Córdoba}}, \bibinfo {author} {\bibfnamefont {D.}~\bibnamefont {Jaksch}}, \ and\ \bibinfo {author} {\bibfnamefont {F.}~\bibnamefont {Schlawin}},\ }\href {http://arxiv.org/abs/2407.08405} {{\selectlanguage {en}\enquote {\bibinfo {title} {Floquet {Schrieffer}-{Wolff} transform based on {Sylvester} equations},}\ }} (\bibinfo {year} {2024}),\ \bibinfo {note} {arXiv:2407.08405 [cond-mat, physics:quant-ph]}\BibitemShut {NoStop}%
\bibitem [{\citenamefont {Starkov}\ \emph {et~al.}(2023)\citenamefont {Starkov}, \citenamefont {Fistul},\ and\ \citenamefont {Eremin}}]{starkov2023}%
  \BibitemOpen
  \bibfield  {author} {\bibinfo {author} {\bibfnamefont {G.~A.}\ \bibnamefont {Starkov}}, \bibinfo {author} {\bibfnamefont {M.~V.}\ \bibnamefont {Fistul}}, \ and\ \bibinfo {author} {\bibfnamefont {I.~M.}\ \bibnamefont {Eremin}},\ }\href {\doibase 10.1103/PhysRevB.108.235417} {\bibfield  {journal} {\bibinfo  {journal} {Physical Review B}\ }\textbf {\bibinfo {volume} {108}},\ \bibinfo {pages} {235417} (\bibinfo {year} {2023})},\ \bibinfo {note} {arXiv:2309.09829 [cond-mat, physics:quant-ph]}\BibitemShut {NoStop}%
\bibitem [{\citenamefont {Wurtz}\ \emph {et~al.}(2020)\citenamefont {Wurtz}, \citenamefont {Claeys},\ and\ \citenamefont {Polkovnikov}}]{wurtz2020}%
  \BibitemOpen
  \bibfield  {author} {\bibinfo {author} {\bibfnamefont {J.}~\bibnamefont {Wurtz}}, \bibinfo {author} {\bibfnamefont {P.~W.}\ \bibnamefont {Claeys}}, \ and\ \bibinfo {author} {\bibfnamefont {A.}~\bibnamefont {Polkovnikov}},\ }\href {\doibase 10.1103/PhysRevB.101.014302} {\bibfield  {journal} {\bibinfo  {journal} {Physical Review B}\ }\textbf {\bibinfo {volume} {101}},\ \bibinfo {pages} {014302} (\bibinfo {year} {2020})}\BibitemShut {NoStop}%
\bibitem [{\citenamefont {Zhang}\ \emph {et~al.}(2022)\citenamefont {Zhang}, \citenamefont {Yang}, \citenamefont {Xu},\ and\ \citenamefont {Li}}]{zhang2022}%
  \BibitemOpen
  \bibfield  {author} {\bibinfo {author} {\bibfnamefont {Z.}~\bibnamefont {Zhang}}, \bibinfo {author} {\bibfnamefont {Y.}~\bibnamefont {Yang}}, \bibinfo {author} {\bibfnamefont {X.}~\bibnamefont {Xu}}, \ and\ \bibinfo {author} {\bibfnamefont {Y.}~\bibnamefont {Li}},\ }\href {\doibase 10.1103/PhysRevResearch.4.043023} {\bibfield  {journal} {\bibinfo  {journal} {Physical Review Research}\ }\textbf {\bibinfo {volume} {4}},\ \bibinfo {pages} {043023} (\bibinfo {year} {2022})}\BibitemShut {NoStop}%
\bibitem [{\citenamefont {Haq}\ \emph {et~al.}(2022)\citenamefont {Haq}, \citenamefont {Iqbal}, \citenamefont {Illahi}, \citenamefont {Ahmad},\ and\ \citenamefont {Nazama}}]{haq2022}%
  \BibitemOpen
  \bibfield  {author} {\bibinfo {author} {\bibfnamefont {R.~U.}\ \bibnamefont {Haq}}, \bibinfo {author} {\bibfnamefont {B.}~\bibnamefont {Iqbal}}, \bibinfo {author} {\bibfnamefont {M.}~\bibnamefont {Illahi}}, \bibinfo {author} {\bibfnamefont {B.}~\bibnamefont {Ahmad}}, \ and\ \bibinfo {author} {\bibnamefont {Nazama}},\ }\href {http://arxiv.org/abs/2208.04746} {{\selectlanguage {en}\enquote {\bibinfo {title} {Schrieffer-{Wolff} {Transformation} on {IBM} {Quantum} {Computer}},}\ }} (\bibinfo {year} {2022}),\ \bibinfo {note} {arXiv:2208.04746 [quant-ph]}\BibitemShut {NoStop}%
\bibitem [{\citenamefont {Mozgunov}(2023)}]{mozgunov2023}%
  \BibitemOpen
  \bibfield  {author} {\bibinfo {author} {\bibfnamefont {E.}~\bibnamefont {Mozgunov}},\ }\href {http://arxiv.org/abs/2302.02458} {{\selectlanguage {en}\enquote {\bibinfo {title} {Precision of quantum simulation of all-to-all coupling in a local architecture},}\ }} (\bibinfo {year} {2023}),\ \bibinfo {note} {arXiv:2302.02458 [quant-ph]}\BibitemShut {NoStop}%
\bibitem [{\citenamefont {Marécat}\ \emph {et~al.}(2023)\citenamefont {Marécat}, \citenamefont {Senjean},\ and\ \citenamefont {Saubanère}}]{marecat2023}%
  \BibitemOpen
  \bibfield  {author} {\bibinfo {author} {\bibfnamefont {Q.}~\bibnamefont {Marécat}}, \bibinfo {author} {\bibfnamefont {B.}~\bibnamefont {Senjean}}, \ and\ \bibinfo {author} {\bibfnamefont {M.}~\bibnamefont {Saubanère}},\ }\href {\doibase 10.1103/PhysRevB.107.155110} {\bibfield  {journal} {\bibinfo  {journal} {Physical Review B}\ }\textbf {\bibinfo {volume} {107}},\ \bibinfo {pages} {155110} (\bibinfo {year} {2023})},\ \bibinfo {note} {arXiv:2212.11089 [cond-mat, physics:quant-ph]}\BibitemShut {NoStop}%
\bibitem [{\citenamefont {Pintér}\ \emph {et~al.}(2024)\citenamefont {Pintér}, \citenamefont {Frank}, \citenamefont {Varjas},\ and\ \citenamefont {Pályi}}]{pinter2024}%
  \BibitemOpen
  \bibfield  {author} {\bibinfo {author} {\bibfnamefont {G.}~\bibnamefont {Pintér}}, \bibinfo {author} {\bibfnamefont {G.}~\bibnamefont {Frank}}, \bibinfo {author} {\bibfnamefont {D.}~\bibnamefont {Varjas}}, \ and\ \bibinfo {author} {\bibfnamefont {A.}~\bibnamefont {Pályi}},\ }\href {http://arxiv.org/abs/2407.10478} {{\selectlanguage {en}\enquote {\bibinfo {title} {The geometry of the {Hermitian} matrix space and the {Schrieffer}--{Wolff} transformation},}\ }} (\bibinfo {year} {2024}),\ \bibinfo {note} {arXiv:2407.10478 [cond-mat, physics:quant-ph]}\BibitemShut {NoStop}%
\bibitem [{\citenamefont {Kowalski}\ and\ \citenamefont {Bauman}(2022)}]{kowalski2022}%
  \BibitemOpen
  \bibfield  {author} {\bibinfo {author} {\bibfnamefont {K.}~\bibnamefont {Kowalski}}\ and\ \bibinfo {author} {\bibfnamefont {N.~P.}\ \bibnamefont {Bauman}},\ }\href {http://arxiv.org/abs/2211.10529} {{\selectlanguage {en}\enquote {\bibinfo {title} {Fock-space {Schrieffer}--{Wolff} transformation: classically-assisted rank-reduced quantum phase estimation algorithm},}\ }} (\bibinfo {year} {2022}),\ \bibinfo {note} {arXiv:2211.10529 [quant-ph]}\BibitemShut {NoStop}%
\bibitem [{\citenamefont {Breuer}\ and\ \citenamefont {Petruccione}(2007)}]{breuer2007}%
  \BibitemOpen
  \bibfield  {author} {\bibinfo {author} {\bibfnamefont {H.~P.}\ \bibnamefont {Breuer}}\ and\ \bibinfo {author} {\bibfnamefont {F.}~\bibnamefont {Petruccione}},\ }\href@noop {} {\emph {\bibinfo {title} {The {Theory} of {Open} {Quantum} {Systems}}}}\ (\bibinfo  {publisher} {Oxford University Press, USA},\ \bibinfo {year} {2007})\BibitemShut {NoStop}%
\bibitem [{\citenamefont {Gardiner}\ and\ \citenamefont {Zoller}(2004)}]{gardiner2004}%
  \BibitemOpen
  \bibfield  {author} {\bibinfo {author} {\bibfnamefont {C.}~\bibnamefont {Gardiner}}\ and\ \bibinfo {author} {\bibfnamefont {P.}~\bibnamefont {Zoller}},\ }\href@noop {} {\emph {\bibinfo {title} {Quantum noise}}}\ (\bibinfo  {publisher} {Springer},\ \bibinfo {year} {2004})\BibitemShut {NoStop}%
\bibitem [{\citenamefont {Roccati}\ and\ \citenamefont {Cilluffo}(2024)}]{roccati2024}%
  \BibitemOpen
  \bibfield  {author} {\bibinfo {author} {\bibfnamefont {F.}~\bibnamefont {Roccati}}\ and\ \bibinfo {author} {\bibfnamefont {D.}~\bibnamefont {Cilluffo}},\ }\href {http://arxiv.org/abs/2402.15556} {{\selectlanguage {en}\enquote {\bibinfo {title} {Controlling markovianity with chiral giant atoms},}\ }} (\bibinfo {year} {2024}),\ \bibinfo {note} {arXiv:2402.15556 [cond-mat, physics:physics, physics:quant-ph]}\BibitemShut {NoStop}%
\bibitem [{\citenamefont {MacDonald}\ \emph {et~al.}(1988)\citenamefont {MacDonald}, \citenamefont {Girvin},\ and\ \citenamefont {Yoshioka}}]{macdonald1988}%
  \BibitemOpen
  \bibfield  {author} {\bibinfo {author} {\bibfnamefont {A.~H.}\ \bibnamefont {MacDonald}}, \bibinfo {author} {\bibfnamefont {S.~M.}\ \bibnamefont {Girvin}}, \ and\ \bibinfo {author} {\bibfnamefont {D.}~\bibnamefont {Yoshioka}},\ }\href {\doibase 10.1103/PhysRevB.37.9753} {\bibfield  {journal} {\bibinfo  {journal} {Physical Review B}\ }\textbf {\bibinfo {volume} {37}},\ \bibinfo {pages} {9753} (\bibinfo {year} {1988})}\BibitemShut {NoStop}%
\bibitem [{\citenamefont {Fazekas}()}]{fazekas}%
  \BibitemOpen
  \bibfield  {author} {\bibinfo {author} {\bibfnamefont {P.}~\bibnamefont {Fazekas}},\ }\href@noop {} {\emph {\bibinfo {title} {Lecture notes on electron correlation and magnetism}}}\BibitemShut {NoStop}%
\bibitem [{\citenamefont {Roccati}\ \emph {et~al.}(2024)\citenamefont {Roccati}, \citenamefont {Bello}, \citenamefont {Gong}, \citenamefont {Ueda}, \citenamefont {Ciccarello}, \citenamefont {Chenu},\ and\ \citenamefont {Carollo}}]{roccati2024a}%
  \BibitemOpen
  \bibfield  {author} {\bibinfo {author} {\bibfnamefont {F.}~\bibnamefont {Roccati}}, \bibinfo {author} {\bibfnamefont {M.}~\bibnamefont {Bello}}, \bibinfo {author} {\bibfnamefont {Z.}~\bibnamefont {Gong}}, \bibinfo {author} {\bibfnamefont {M.}~\bibnamefont {Ueda}}, \bibinfo {author} {\bibfnamefont {F.}~\bibnamefont {Ciccarello}}, \bibinfo {author} {\bibfnamefont {A.}~\bibnamefont {Chenu}}, \ and\ \bibinfo {author} {\bibfnamefont {A.}~\bibnamefont {Carollo}},\ }\href {\doibase 10.1038/s41467-024-46471-w} {\bibfield  {journal} {\bibinfo  {journal} {Nature Communications}\ }\textbf {\bibinfo {volume} {15}},\ \bibinfo {pages} {2400} (\bibinfo {year} {2024})}\BibitemShut {NoStop}%
\end{thebibliography}%

\end{document}